\documentclass[aps,prd, preprint, onecolumn,superscriptaddress,amssymb,amsmath,nofootinbib]{revtex4-1}
\pdfoutput=1

\usepackage{color}
\usepackage{epsfig}
\usepackage{graphicx}
\usepackage{float}
\usepackage{epstopdf}
\usepackage{hyperref} 
\usepackage[usenames,dvipsnames]{xcolor}
\usepackage{slashed}
\usepackage{braket}
\begin{document}

\preprint{EFI-20-11}

\title{Searching for the Higgsino-Bino Sector at the LHC}

\author{Jia Liu}
\email{liuj1@uchicago.edu}
\affiliation{Physics Department and Enrico Fermi Institute, University of Chicago, Chicago, IL 60637}

\author{Navin McGinnis}
\email{nmmcginn@indiana.edu}
\affiliation{Physics Department, Indiana University, Bloomington, IN 47405, USA}
\affiliation{High Energy Physics Division, Argonne National Laboratory, Argonne, IL, 60439}

\author{Carlos E.M. Wagner}
\email{cwagner@anl.gov}
\affiliation{Physics Department and Enrico Fermi Institute, University of Chicago, Chicago, IL 60637}
\affiliation{High Energy Physics Division, Argonne National Laboratory, Argonne, IL, 60439}
\affiliation{Kavli Institute for Cosmological Physics, University of Chicago, Chicago, IL, 60637}

\author{Xiao-Ping Wang}
\email{xia.wang@anl.gov}
\affiliation{High Energy Physics Division, Argonne National Laboratory, Argonne, IL, 60439}

        %\thanks{}
%\altaffiliation {}

%\date{\today}
\date{\today}

\begin{abstract}
We study the  search for electroweakinos at the 13 TeV  LHC in the case of heavy scalar superpartners. We consider
both the direct production mode and the one associated with the decay of heavy Higgs bosons, and concentrate
on the case of light Higgsinos and Binos. In this case, the direct production searches becomes more challenging
than in the light Wino scenario. In the direct production mode, we use the current experimental searches to
set the reach for these particles at larger luminosities, and we emphasize the relevance of considering both
the neutral gauge boson and the neutral Higgs decay modes of the second and third lightest neutralino. 
We show the complementarity of these searches with the ones induced by the decay of the heavy Higgs
bosons, which are dominated by the associated production of the lightest neutralino with the second
and third lightest ones, with the latter decaying into gauge bosons.  We show that, depending on the value of 
$\tan\beta$, the Higgs boson decay channel remains competitive with the direct production channel up to heavy 
Higgs boson masses of about 1~TeV. Moreover, this search is not limited by the same kinematic considerations 
as the ones in the direct production mode and can cover masses up to the kinematic threshold for the decay of the heavier electroweakinos into the lightest neutralino.  This
decay mode provides also an alternative way of looking 
for heavy Higgs bosons in this range of masses and hence should be a high priority for future LHC analyses. 
\end{abstract}


\maketitle

\tableofcontents
%\newpage

%\section{ \label{sec:}}

%\subsection{}

%\subsubsection{}

\section{Introduction}
The Standard Model provides an excellent low energy effective theory 
describing the interaction of particles at energies similar to or smaller than the weak scale.  Searches at the LHC are
underway looking for the possible production of new particles with masses up to a few TeV. No clear 
signal of new physics has been found, leading to strong bounds on the presence of strongly interacting particles
at the TeV scale~(see, for instance, Refs.~\cite{Aaboud:2017vwy, Sirunyan:2018vjp}).  
However, if the particles are weakly interacting, new physics searches may be hindered by small 
production cross sections and large backgrounds induced by the production of the weak gauge bosons
and strongly interacting states, like the top and bottom quarks.  Contrary to strongly interacting 
states,  many weakly interacting particle search channels will become statistically significant
 only at high luminosities.  For this reason,  the LHC experimental collaborations are now starting to put stronger
 emphasis on the search  for weakly interacting particles \cite{Sirunyan:2017zss, Sirunyan:2017lae, Sirunyan:2018ubx, Sirunyan:2018iwl, Aaboud:2018jiw, Aaboud:2018sua, Aaboud:2018ngk, Aad:2019vvf, ATLAS-CONF-2019-008, ATLAS-CONF-2019-020, Aad:2020qnn}. 
 
 Among the many hypothetical weakly interacting particles, the electroweakinos, namely the superpartners
 of the Higgs and electroweak gauge bosons, are well motivated examples \cite{Fayet:1976et, Fayet:1977yc, Gunion:1987kg}.  They
 are a natural consequence of supersymmetry at the TeV scale \cite{Haber:1984rc,Martin:1997ns,Barbieri:1987fn, deCarlos:1993rbr,Brust:2011tb}.  
 Weakly interacting supersymmetric particles
 are not subject to strong renormalization group effects in the evolution of their masses from high scales,
 and are therefore expected to be lighter than strongly interacting ones.  Moreover, this sector includes  the 
 Dark Matter candidates within these theories \cite{Goldberg:1983nd, Ellis:1983ew}. Since Dark Matter is arguably one of the
 strongest reasons to expect new physics among the Standard Model, the search for these particles
 is particularly well justified. The superpartner of the hypercharge gauge boson, the Bino,  places a special 
 role beyond these particles, since it does not couple to any of the electroweak gauge bosons and  
 therefore, after mixing with the heavier superpartners of the Higgs and the electroweak gauge bosons,
 can acquire the proper relic density for masses significantly lower than the TeV scale. The superpartners
 of the Higgs bosons, the Higgsinos,  are very relevant in defining the light Dark Matter candidate interactions,
 since they mix with the Binos through the Higgs vacuum expectation values. In
 this work, we shall assume, for simplicity, that the superpartners of the weak gauge bosons, the Winos,
 are significantly heavier than the two neutral Higgsino states, which therefore will become the second and
 third lightest neutralinos.  Moreover, we shall assume that all scalar superpartners
 are sufficiently heavy, so that  they do not participate in a relevant way in the decays of the Higgs bosons or electroweakinos.   
  
Searches for electroweakinos at the LHC have so far been performed in direct production channels
\cite{Aad:2019vvf, Aaboud:2018ngk, Aad:2020qnn, ATLAS-CONF-2019-008, Aaboud:2018sua, ATLAS-CONF-2019-020, Sirunyan:2018ubx, Sirunyan:2018iwl, Sirunyan:2017zss, Sirunyan:2017lae}. 
The main production channel is mediated by charged and neutral electroweak gauge bosons, and
can lead to hadronic, semi-leptonic and purely leptonic final states plus missing energy.  Of particular 
relevance among these channels is the associated production of charged and neutral electroweakino
states, with charged states decaying into  $W^\pm$ bosons and the lightest neutralino state, and
the second (and third) lightest neutralino states decaying into neutral gauge and Higgs bosons. 
The assumption of heavy Winos and scalar superpartners tend to weaken the reach of the LHC,
since Winos have higher cross sections than the Higgsinos and light sleptons tend to contribute
to the electroweakino decays increasing the total branching ratio into leptons, which are easier to 
distinguish from the large QCD background.   Since the LHC limits are usually presented considering
Wino direct production, e.g.~\cite{Aad:2019vvf}, we shall recast these searches by studying the Higgsino cross sections and the proper decay branching ratios into SM Higgs and $Z$-boson final states. 
 
Although well motivated,  searches for electroweakinos proceeding from the decay of heavy Higgs bosons 
are in a preliminary state.   There has  so far been only a few relevant theoretical analyses of heavy Higgs
bosons decaying into electroweakino states \cite{Moortgat:2001pp, Denegri:2001pn, Moortgat:2001wj, Ball:2007zza, Arganda:2012qp, Han:2013gba, Baer:2013xua, Baer:2015tva, Craig:2015jba, Barman:2016kgt, Kulkarni:2017xtf, Duan:2017ucw, Medina:2017bke, Arganda:2018hdn, Bahl:2018zmf, 
Gori:2018pmk, Bahl:2019ago, Alipour-fard:2018mre, Canepa:2020ntc, Adhikary:2020ujn}, concentrating mostly on the associated production of the
lightest, and second and third lightest neutralinos, which then lead mostly to decay channels involving single production of the $Z$
and Higgs bosons plus missing energy. Similar studies have been performed in the Next-to-Minimal-Supersymmetric-Standard-Model 
(NMSSM)~\cite{Christensen:2013dra, Dutta:2014hma, Han:2014nba, Wang:2015omi, Ellwanger:2017skc, Baum:2017gbj, Baum:2019uzg}. 
These analyses show the relevance and the promising 
reach of these channels and serve as motivation for our work. For the comparison with the direct
production channel, it is most relevant to understand the reach in the electroweakino mass parameter
space, something not provided by these analyses. 

In this work, we shall analyze the complementarity of the direct production and Higgs boson decay
channels. The work is organized as follows : 
Section~\ref{sec:electroweakino-crosssections} provides an analysis of the electroweakino 
and Higgs production cross sections. Section~\ref{sec:direct-search-recast} provides the recast
of existing searches to higgsino-like electroweakinos.
In section~\ref{sec:search-from-heavy-Higgs}, we concentrate on the  di-lepton plus missing energy 
channel for electroweakinos from the decay of heavy neutral Higgses and then present the complementarity of this search channel  with the projected bounds for direct production.
Section~\ref{sec:conclusion} is reserved for our conclusions.

\section{Electroweakino productions}
Scenarios where all strongly-interacting superpartners are decoupled from the rest of the MSSM are particularly well motivated both from the current status of direct searches at the LHC and the measurement of the Higgs boson mass. Indeed, within the MSSM, a 125~GeV SM-like Higgs boson mass may only be obtained for masses of the superpartners of the top quark of the order of 2~TeV or larger~\cite{Draper:2013oza,Bagnaschi:2014rsa,Vega:2015fna,Lee:2015uza,Bahl:2017aev}.
Disregarding sleptons for the moment, the remaining states of the MSSM then are the superpartners of the weak gauge bosons and Higgs doublets, and additional Higgs bosons. Assuming this low-energy spectrum leads to a particularly simple set of states, whose dynamics are determined by only a few parameters. The neutralinos appear as mixtures of the Bino, Wino, and the Higgsinos whose mass matrix is given by

\begin{equation}
M_{N}=\begin{pmatrix}
M_{1} & 0 & -c_{\beta}s_{W}m_{Z} & s_{\beta}s_{W}m_{Z}\\
0 & M_{2} &c_{\beta}c_{W}m_{Z} & -s_{\beta}c_{W}m_{Z}\\
-c_{\beta}s_{W}m_{Z} & c_{\beta}c_{W}m_{Z} & 0 & -\mu\\
s_{\beta}s_{W}m_{Z} & - s_{\beta}c_{W}m_{Z} & -\mu & 0
\end{pmatrix},
\end{equation}
where $c_{W}=\cos\theta_{W}$, $s_{W}=\sin\theta_{W}$, and $\theta_{W}$ is the weak-mixing angle, $c_{\beta}=\cos\beta$, $s_{\beta}=\sin\beta$, and $\tan\beta = \braket{H_{u}^{0}}/\braket{H_{d}^{0}}$ is the ratio of the vacuum expectation values of the Higgs doublets. The mass $M_1$ is the Bino mass parameter, $M_2$  and $\mu$ are the Wino and Higgsino mass parameters, respectively.
Similarly, the chargino states appear as mixtures of charged Wino and Higgsino components with mass matrix

\begin{equation}
M_{C}=\begin{pmatrix}
M_{2} & \sqrt{2}s_{\beta}m_{W}\\
 \sqrt{2}c_{\beta}m_{W} & \mu
\end{pmatrix}.
\end{equation}
For a more detailed review of the resulting mass eigenstates and couplings see Ref.~\cite{Martin:1997ns} and references therein. Including the Higgs sector of the MSSM, the low energy theory results in a rich phenomenology where the relevant parameters are $\{M_{1}, M_{2}, \mu, \tan\beta\}$ and the masses of the heavy Higgs bosons, which are characterized with the CP-odd Higgs mass parameter $m_{A}$. In the case when the Wino is also decoupled the lightest electroweakinos are Bino-, and Higgsino-like. In this limit, the results we present in the following sections relies mainly on only four parameters $\{M_{1}, \mu, \tan\beta, M_{A}\}$.

\label{sec:electroweakino-crosssections}
\subsection{Direct Electroweakino Production}

The direct production of electroweakino states is governed by the interchange of $W^{\pm}$ and
$Z$ gauge bosons (see for instance, Ref.~\cite{Martin:1997ns}). Since the Bino does not couple to the gauge bosons, all production modes
are controlled by the Higgsino couplings to gauge bosons, and the production of a particular
mass eigenstate will depend on its component of the Higgsino states.  Assuming the 
lightest supersymmetric particle is mostly a Bino, its production, although kinematically favored,  may 
only happen through its mixing with the neutral Higgsino states. Pair production of Bino
states may be measured via its recoil against jets, in the jets plus missing energy channel.
However, the production cross section tends to be too small to be competitive with other
search channels.  Binos may be hence mostly produced in association with the
charged and neutral Higgsino states, leading after the decay of the heavier states to
final states consisting of $h$, $Z$ or $W^\pm$ plus missing energy.   However, unless the
Bino and Higgsino masses are close to each other, the cross sections are again too
small to make these competitive search channels. 

Much more promising is the search for electroweakinos via the production of the heavier,
but mainly Higgsino states.  Since it is not hampered by mixing, this search is only 
limited by the kinematic effects related to low production cross sections associated with
high mass states. Also, in the limit of heavy superpartners, the Higgsino cross sections
are smaller than the Wino cross sections, implying also a smaller reach than in the case
of light Winos.  
Interestingly enough, since the scalar leptons couple to the Higgsino
states in a way proportional to Yukawa couplings, the search for Higgsinos tends to be
minimally affected by the presence of such states, which in this scenario will therefore
mainly decay into leptons and missing energy and can be searched separately from
the Higgsino states. This is relevant, since light sleptons may contribute to an explanation 
of the $g-2$ anomaly~\cite{Ellis:1982by,Moroi:1995yh,Carena:1996qa,Czarnecki:2001pv,Feng:2001tr,Martin:2001st} . 
Observe, however, that the contributions to the muon anomalous magnetic
moment are highly reduced in the case of heavy Winos.
In fact, even if the slepton masses were only a few hundreds of GeV, the contributions to $(g-2)_\mu$ in the region of parameters 
we are considering, for moderate values of $\mu$ and $\tan\beta$, is  only a few times $10^{-10}$ and not large enough  to
fit the Brookhaven experiment observation~\cite{Bennett:2006fi}, 
\begin{equation}
\delta a_\mu^{\rm exp} \simeq \left( 27 \pm 7 \pm 5   \right) \times 10^{-10},
\end{equation} 
where the errors are associated with experimental and theoretical uncertainties.  Fitting the Brookhaven experiments
would demand large values of $\tan\beta$  pushing the Higgs masses to values larger than 1~TeV~\cite{Sirunyan:2018zut, Aad:2020zxo}, 
for which the Higgs boson decay into electroweakinos ceases to be a competitive search channel.

Regarding the Higgsino being a Dark Matter candidate, if one assumes that the Dark Matter  relic density comes from thermal annihilation,
one needs to mix Higgsino and Bino appropriately to reach the right relic abundance. This is the so-called well tempered neutralino 
region~\cite{ArkaniHamed:2006mb}, which tends to be in tension with direct detection experiments~\cite{Akerib:2016vxi, Cui:2017nnn, Aprile:2017iyp, Aprile:2018dbl, Baer:2016ucr,Huang:2017kdh}. 
However, the relic density constraints can be easily evaded in the case of light Binos by the assumption of late entropy production (see, for instance, Ref.~\cite{Gelmini:2006pq}). Other ways to avoid the relic density constraints in the region of parameters we are concentrating on include coannihilation, $Z$-resonance annihilation , $h$-resonance annihilation   (see, for instance, Refs.~ \cite{Ellis:1999mm,Han:2016qtc,Carena:2018nlf,Carena:2019pwq} ).  Moreover, it is easy to weaken
the  spin-independent direct Dark Matter search constraints by simply concentrating on negative values of $\mu$~(see Refs.~\cite{Baer:2008ih, Huang:2014xua, Huang:2017kdh, Abdughani:2017dqs, Baum:2017enm, Endo:2017zrj, Han:2018gej, Ellis:2000ds}). Though, it is worth noting that the constraints from spin-dependent Dark Matter searches will still apply~\cite{Bagnaschi:2017tru, Cao:2019qng}.

\begin{figure}[htb]
\centering
	 \includegraphics[scale=0.5]{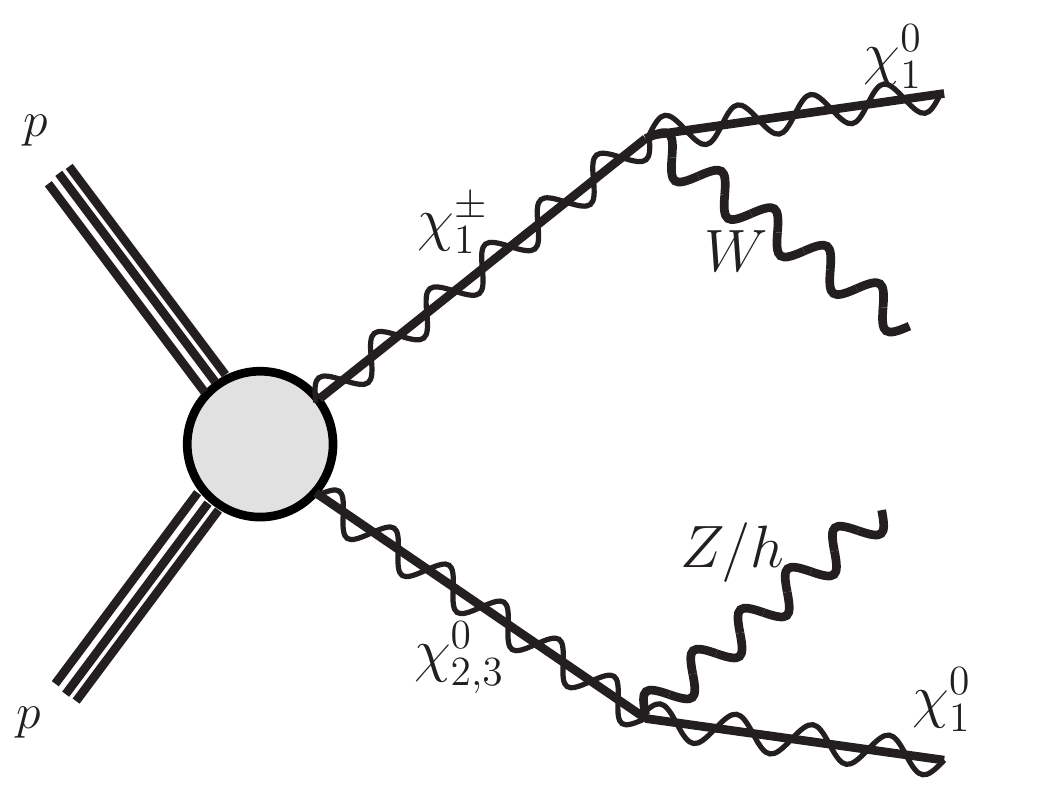} \hspace{1cm}
	 \includegraphics[scale=0.5]{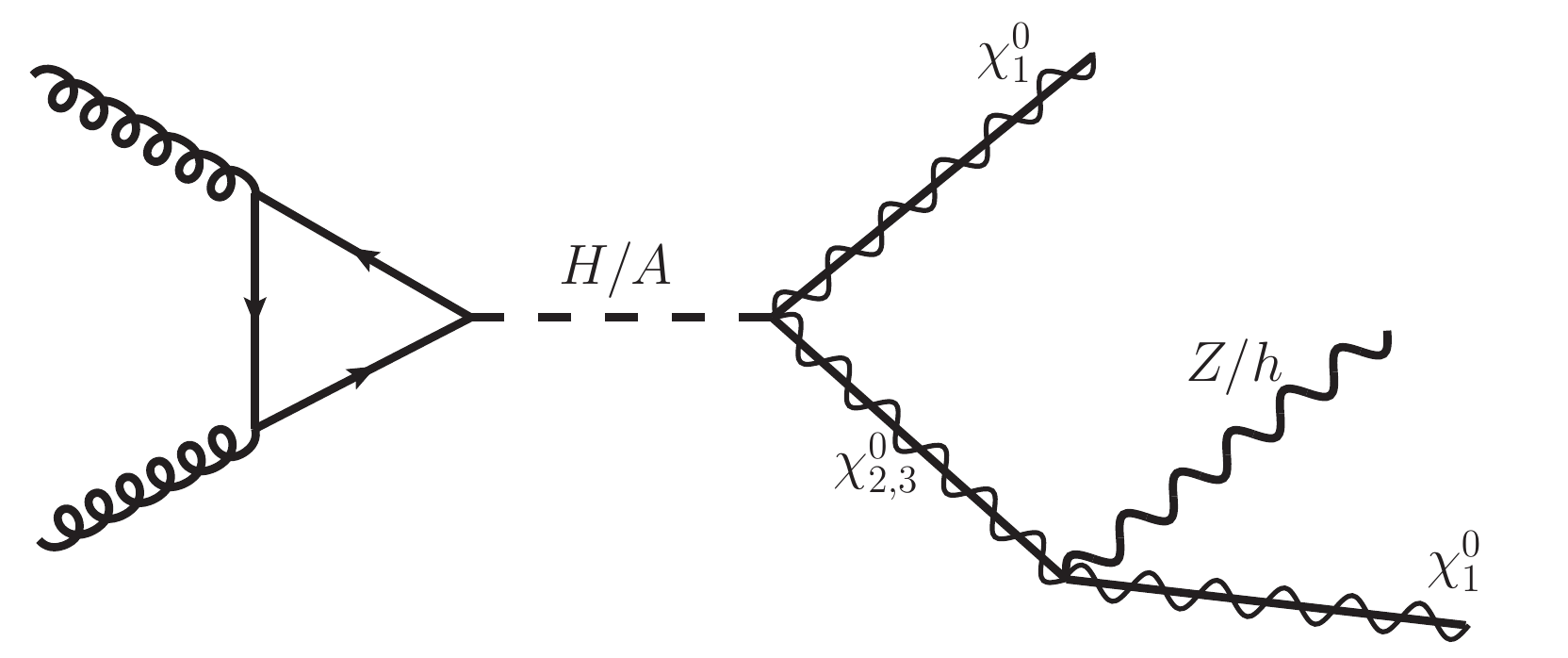}
\caption{Representative decay topologies of electroweakino production at the LHC. {\bf Left:} Direct production channels resulting from pair producing a chargino and neutralino. {\bf Right:} The Higgs portal channel resulting from the production of a heavy neutral Higgs boson decaying to neutralino pairs.}
\label{fig:diagrams}
\end{figure}

For the reasons discussed above, we will be mostly interested in the production
of mainly Higgsino states with negative values of $\mu$.  The spectrum will consist of two neutral Majorana states,
the second and third lightest neutralinos, and one charged Dirac state.  The associated
direct production of the charged and neutral states, mediated mostly by the charged gauge boson $W^\pm$ 
carries the largest cross section and it is represented in Fig.~\ref{fig:diagrams}. Since the 
second and third lightest neutralinos are
close in mass, provided it is kinematically allowed, they will mostly decay into either
$Z$ or light Higgs $h$ final states plus missing energy. The chargino state, instead
will decay into $W^\pm$ and missing energy
\begin{flalign}
pp\rightarrow \chi_{1}^{\pm} + \chi_{h}^{0}\rightarrow W^{\pm} + Z/h + \slashed{E}_{T},
\end{flalign}
where $h=2,3$.  We will concentrate in the region of parameters where the second and third
lightest neutralino can decay into on-shell $Z$ gauge bosons. Therefore, considering the decay of
the $W^\pm$ into lepton states, the most interesting final states will be either
tri-leptons plus missing energy or pairs of bottom quarks plus one lepton plus missing
energy.  Considering the decay of the $W^\pm$ into hadrons, the most interesting
channel is two leptons (from $Z$) or two bottoms (from $h$) and missing energy.  

\begin{figure}[htb]
\centering
	\includegraphics[scale=0.75]{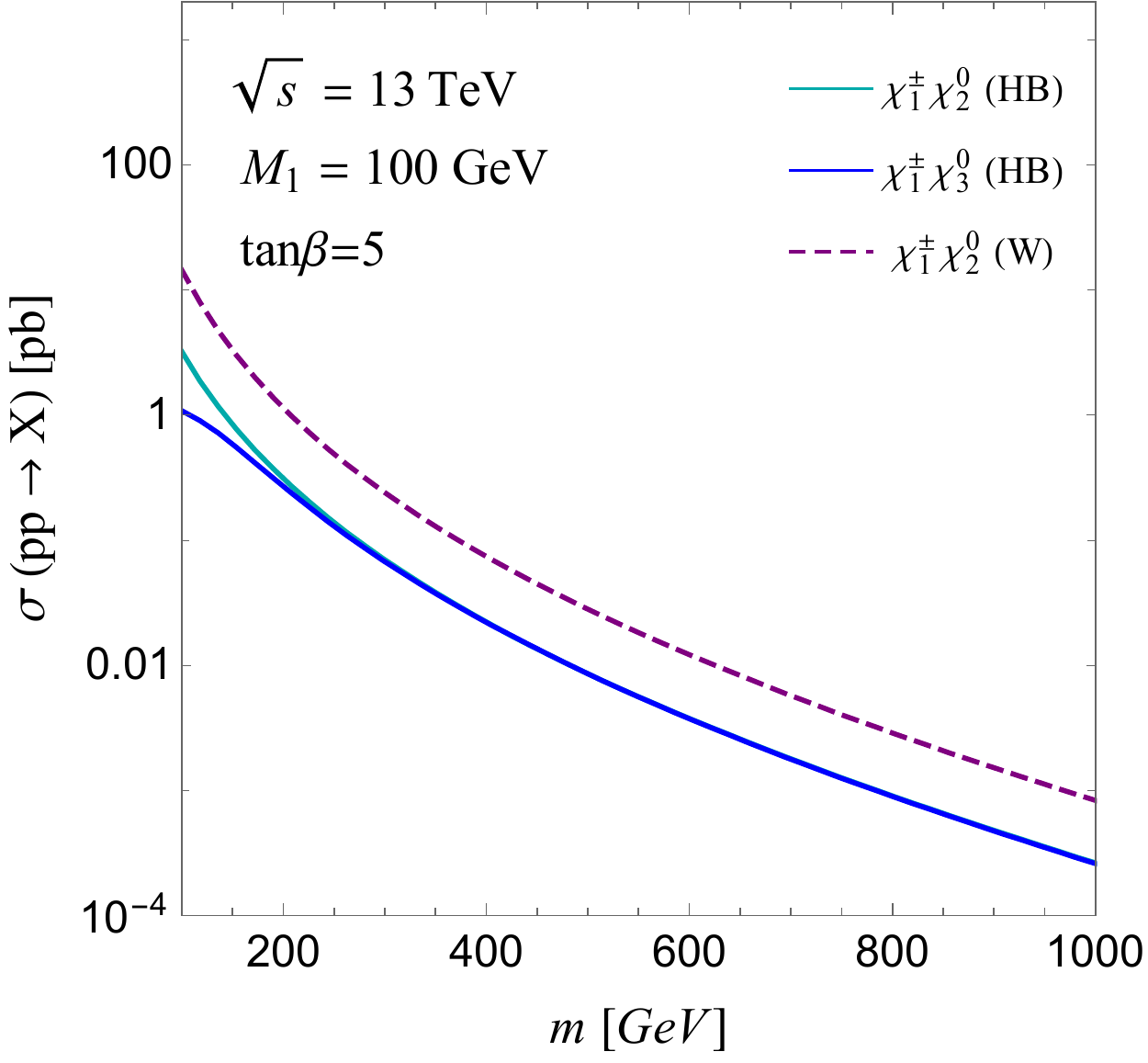} 
\caption{Comparison of the  LO direct production cross sections of electroweakinos in the pure Wino scenario (dashed lines) and the Bino-Higgsino scenario (solid lines) with $M_{1}=100$ GeV and $\tan\beta=5$, assuming that scalar superpartners are decoupled. In the former case, $\mu=-2$ TeV and $m=M_{2}$ is varied. In the latter, $M_{2}=2$ TeV and $m=|\mu|$ is varied.}
\label{fig:ew_prod}
\end{figure}

In Fig.~\ref{fig:ew_prod} we present the dependence of the associated production
cross section as a function of the Higgsino mass parameter $\mu$ that we 
take to be negative for this consideration, and that represents the overall scale
of the Higgsino masses (there is a small dependence of the Higgsino spectrum 
on the sign of $\mu$, which is however, only significant for low values of $\mu$,
of the order of 100~GeV).  We also represent, for comparison, the production
of Winos, as a function of the Wino mass parameter $M_2$, making evident the 
larger cross sections associated with these states. 

\begin{figure}[htb]
\centering
	\includegraphics[scale=0.9]{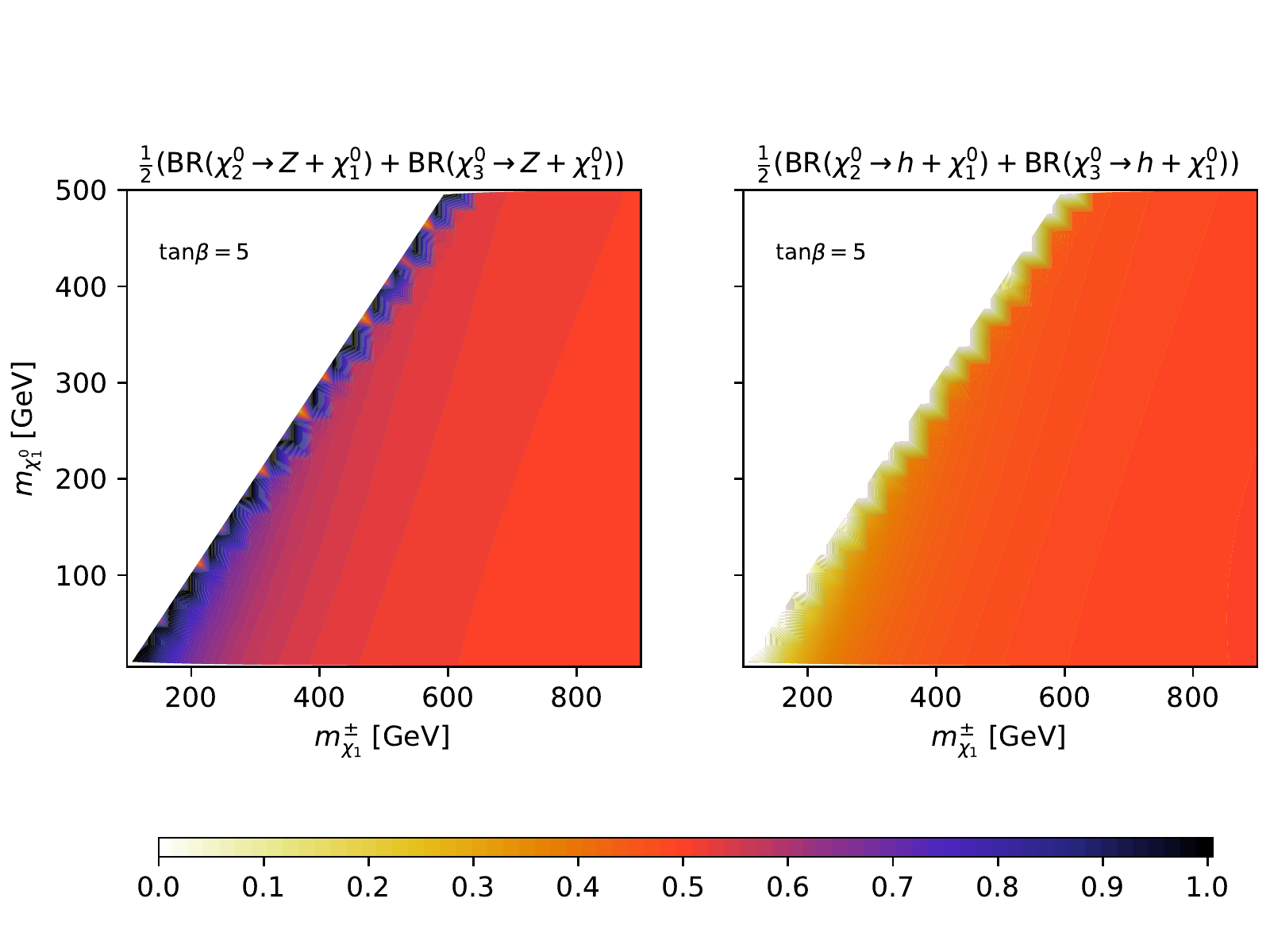}
\caption{{\bf Left:} Sum of the total branching ratio of Higgsinos to the Z boson, $\chi^{0}_{2,3}\rightarrow Z + \chi^{0}_{1}$, normalized by $1/2$. {\bf Right:} Sum of the total branching ratio of Higgsinos to the SM Higgs, $\chi^{0}_{2,3}\rightarrow h + \chi^{0}_{1}$, normalized by $1/2$. We show contours resulting from the scan in Eq.~(\ref{eq:params}), for $\tan\beta = 5$.}
\label{fig:decayZ_h}
\end{figure} 

\iffalse
\begin{figure}[htb]
\centering
	\includegraphics[scale=1]{figures/BR_x23_h_tanb.pdf}
\caption{{\bf Left:} Sum of the $\chi^{0}_{2,3}\rightarrow h + \chi^{0}_{1}$ branching ratios normalized by $1/2$. We show contours resulting from the scan in Eq.~(\ref{eq:params}), for $\tan\beta = 2,5,8,10$.}
\label{fig:decayh}
\end{figure} 
\fi

The branching ratios also play an important role. The presence of two different
Higgsino states imply that their decay  branching ratios will not be equal. 
These two Higgsino states are close in mass, particularly
for masses larger than Higgs mass, 125~GeV, for which also the dependence 
on the sign of $\mu$ becomes less significant.  It is therefore useful to represent
the sum of the branching ratios of the decay of the second and third lightest 
neutralinos into $Z$ and Higgs final states.  These are given in Fig.~\ref{fig:decayZ_h}. In each case, we calculate the branching ratio 
using the code {\tt Spheno} \cite{Porod:2003um,Porod:2011nf} as functions of the parameters

\begin{align}
 M_{1} \in [5,500], \quad
-\mu \in[100,900], \quad
\label{eq:params}
\end{align}
assuming $M_{2} = 2 \text{ TeV}$, $\tan\beta = 5$, and $M_{A} = 1 $ TeV. \footnote{Note that for lower values of the heavy Higgs mass, such that $M_{A} < m_{\chi_1^{\pm}} + m_{\chi_{1}^{0}}$, decay channels of electroweakinos to heavy Higgs bosons may become kinematically open. However, the corresponding branching ratios of these channels would be two small to be of any relevance.}  For large values of $|\mu|$ the two neutralino
Majorana states behave effectively like a single neutral Dirac state and, due
to the Goldstone equivalent theorem, one expects that, approximately, the
neutral states will decay 50~percent of the time into $Z$ and 50 percent
of the time into $h$ final states, something that is evident from 
Fig.~\ref{fig:decayZ_h}.  

In the low mass range, provided both neutralinos become lighter
than the Standard Model-like Higgs boson, they will decay into $Z$ final states 100~percent of
the time. However, we remark that for low values of $\tan\beta$ the splitting can be large enough so that
whenever one neutralino is lighter than the Higgs the other becomes heavier,
up to the point when the lightest neutralino can no longer decay into on-shell
$Z$ bosons. Apart from this, the total branching ratios of electroweakinos to the $Z$ or Higgs boson display the same behavior for values of $\tan\beta$ that we consider.

\subsection{Electroweakinos from heavy Higgs production}

The production of electroweakinos from the decay of heavy Higgs bosons
has a different parametric dependence than the direct production. This
is due to the fact that, contrary to the coupling to $Z$ gauge bosons, pure Higgsino states do not couple to the heavy
Higgs boson due
to gauge invariance, and the coupling always involves mixed Higgsino and
gaugino (Bino) components. Hence, the heavy Higgs decays into a mostly Higgsino
and a mostly Bino state is preferred with respect to both the relevant couplings and kinematics. 

\begin{figure}[h]
\centering
	\includegraphics[width=0.89\columnwidth]{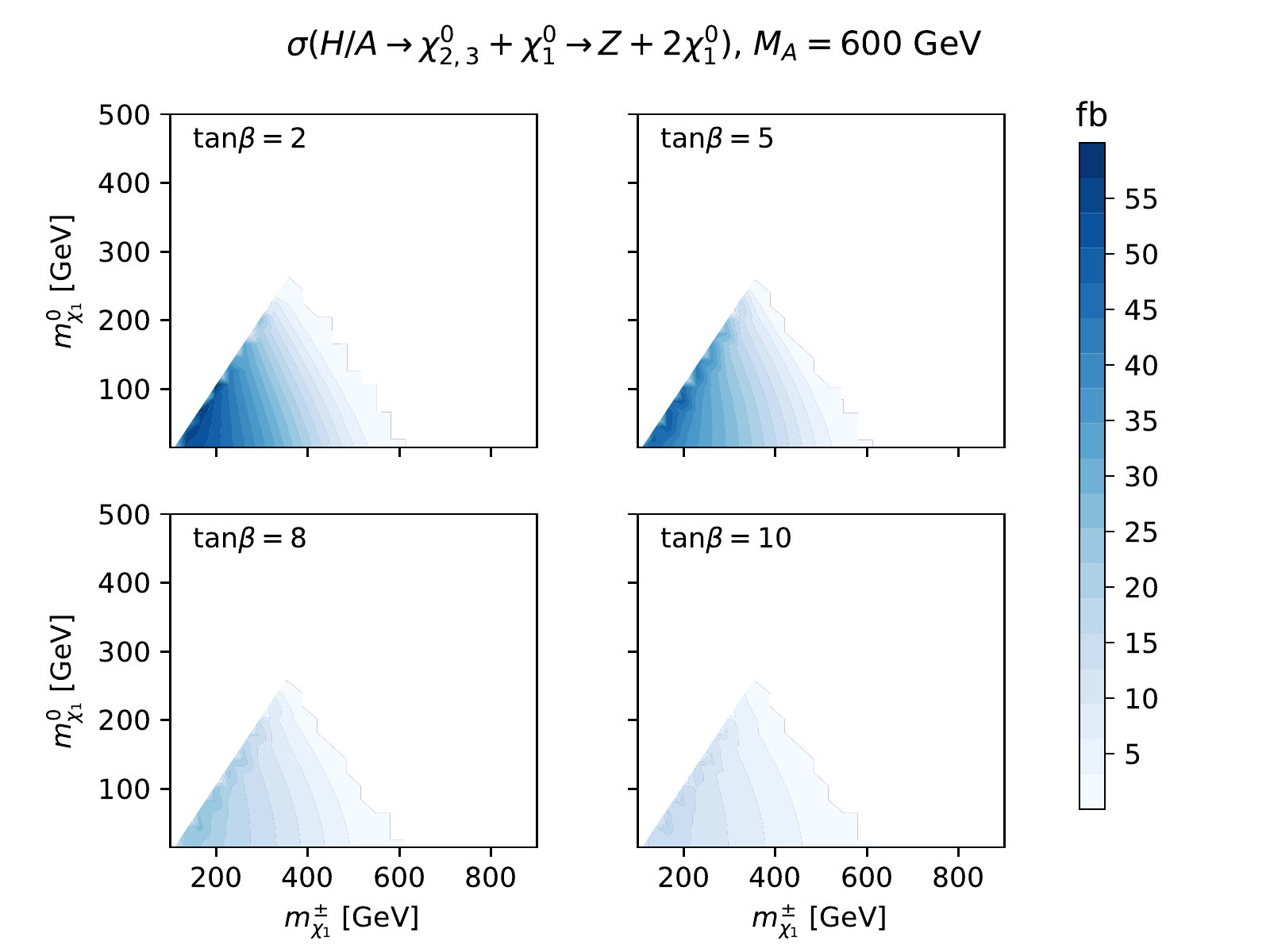} 
\caption{The production cross-section for the process $H/A \to \chi_{2,3}^0 \chi_1^0 \to Z+2\chi_1^0$ for higgsino-like electroweakinos, with heavy scalar mass $m_A = 600$ GeV and $\tan\beta = 2~,5,~8,~10$ respectively.}
\label{fig:higgs_xsection}
\end{figure} 

There are hence two main possibilities. 
Either the lightest neutralino is produced from the decay of neutral
Higgs bosons, in association with the second and third lightest neutralinos, 
or it is produced in association with the chargino states, coming from
the decay of the charged Higgs boson.  In the case of neutral 
Higgs bosons, which is diagrammatically represented in Fig.~\ref{fig:diagrams},
where the second and third lightest neutralino can decay into either a 
$Z$ or a Higgs boson $h$ (the heavier Higgs bosons are highly 
degenerate in mass), leading to interesting mono-Higgs and mono-$Z$
final states.  Among the mono-$Z$ final states, the decay into two 
leptons becomes particularly interesting as has already been
emphasized in Refs.~\cite{Gori:2018pmk}--\cite{Baum:2019uzg}. 

The charged Higgs production mode is also interesting.  Charged 
Higgs, however, can only be produced at sufficiently high rates
in association with top and bottom quarks, implying significantly
smaller cross sections than the ones associated with neutral
gauge bosons. Also, the final state is similar to the top-quark
pair production mode, albeit with significant missing energy.
Hence the reach is hampered by the large $\bar{t}t$ background.
For these reasons, in this work we shall not concentrate on
this mode, but we plan to come back to it in an independent
analysis.

\section{Reach of direct production searches of higgsino states at Higher Luminosities}
\label{sec:direct-search-recast}

In order to compute the reach for Higgsino states in the direct
production mode, we have considered a number of recent ATLAS studies, Ref.~\cite{Aaboud:2018ngk, Aaboud:2018jiw, Aad:2019vvf, ATLAS-CONF-2019-020}, in which they present bounds on the masses of charginos and neutralinos at the 13 TeV LHC for luminosities ranging from $36\text{ fb}^{-1}$ to $139\text{ fb}^{-1}$.\footnote{The CMS study shows similar sensitivities when luminosity is the same, see \cite{Sirunyan:2018ubx}.}
The bounds presented in these studies assume that the electroweakino spectrum results from parameters aligned with the pure Wino scenarios and that the corresponding branching ratios of neutralinos into either the $Z$ boson or SM $h$ are 100~percent, depending on the study. We have recasted
each bound including the cross sections associated with the production of Higgsino-like states, and incorporated the branching ratios
of the neutralino states for a given chargino mass (which is also degenerate
in mass with the neutral Higgsino states for large values of $\mu$) resulting from the scan of Eq.~(\ref{eq:params}).

To obtain the relevant cross section starting from the parameters $M_{1}, \mu$ and $\tan\beta$, mass eigenstates, mixings, and branching ratios are calculated with {\tt FeynHiggs} \cite{Bahl:2018qog,Bahl:2017aev,Bahl:2016brp,Hahn:2013ria,Hollik:2014bua,Degrassi:2002fi,Heinemeyer:1998np,Heinemeyer:1998yj} linked with {\tt SPheno}. The branching ratios of EW-inos to SM gauge bosons have also been checked with {\tt SUSY-HIT} \cite{Djouadi:2006bz}.  When presenting bounds on direct search channels we consider fixed $\tan\beta=5$. In the Higgsino-Bino limit, the $\tan\beta$ dependence on the relevant branching ratios follows as $ \text{BR} (\chi_{2}^{0}\rightarrow Z/h  + \chi_{1}^{0})\sim \frac{1}{2}(1\mp \sin(2\beta))$ and $ \text{BR} (\chi_{3}^{0}\rightarrow Z/h  + \chi_{1}^{0})\sim \frac{1}{2}(1\pm \sin(2\beta))$, when $|\mu| - M_{1}$ is significantly larger than $m_{h}$~\cite{Han:2013kza}. Thus, since we sum both contributions from $\chi_{2}^{0}$ and $\chi_{3}^{0}$ in the production channels, the overall dependence of the total cross is dominated by kinematics rather than the coupling, as expected by the Goldstone Boson Equivalence theorem. In contrast, the Higgs production channel will depend strongly on $\tan\beta$. We will explore this complementarity of the parameter space further in the following section. 

LO cross sections were generated and cross checked between {\tt MadGraph5}  \cite{Alwall:2014hca} and {\tt Prospino2.1} \cite{Beenakker:1996ed}. For chargino masses $\sim 100$ GeV the K-factors between LO and NLO elecroweakino cross sections are about $K\simeq 1.3$  \cite{Beenakker:1999xh}. We find that the K-factors decrease to $K\simeq 1.1$, once the mass of the chargino reaches $\sim 500$ GeV. As this variation occurs in a region which is relevant to the current bounds, we have used the NLO cross sections from {\tt Prospino2.1} in our results. For each chargino-neutralino pair in the scan we then compare the cross section to the corresponding $95\%$ CL upper limit, $ \sigma_{\rm limit}^{\rm up} $, taken from {\tt HEP-Data}

\begin{align}
N_{\rm sig}^{} (N_{\rm bkg}, 95\%)= \sigma_{\rm limit}^{\rm up} \epsilon_{\rm eff} L.
\end{align}

Assuming that the background efficiencies remain the same, we then find the bound for Higgsino-Bino masses via
\begin{align}
\sigma\left(pp \to \chi_1^\pm \chi_{2,3}^0 \right) {\rm BR} \left(\chi_1^\pm  \to W^\pm +\chi_1^0 \right) {\rm BR}\left( \chi_{2,3}^0  \to Z/h  + \chi_1^0 \right)
= \sigma_{\rm limit}^{\rm up}.
\end{align}

Finally, we extrapolate the existing bounds for the $13$ TeV LHC to luminosities of 300 fb$^{-1}$ and 3 ab$^{-1}$
\begin{equation}
 \sigma_{\rm limit}^{\rm up; L^{\prime}} = \sqrt{\frac{L^{exp}}{L^{\prime}}} \sigma_{\rm limit}^{\rm up},
\end{equation}
where $L^{exp}$ is the corresponding luminosity for each analysis that we have considered.

 Inspection of the LHC experiment analyses reveal that the most powerful searches
are associated with the tri-lepton final states and the Higgs final state. In the following subsection, we review the status of the different search channels for Higgsino-like states. We note that since the efficiencies
tend to be improved with further analyses and the energy will be eventually upgraded
to values of order of 14~TeV our analysis should be considered as a conservative one.

\subsection{ZW channels}

\begin{figure}[h!]
\centering
	\includegraphics[width=0.48 \columnwidth]{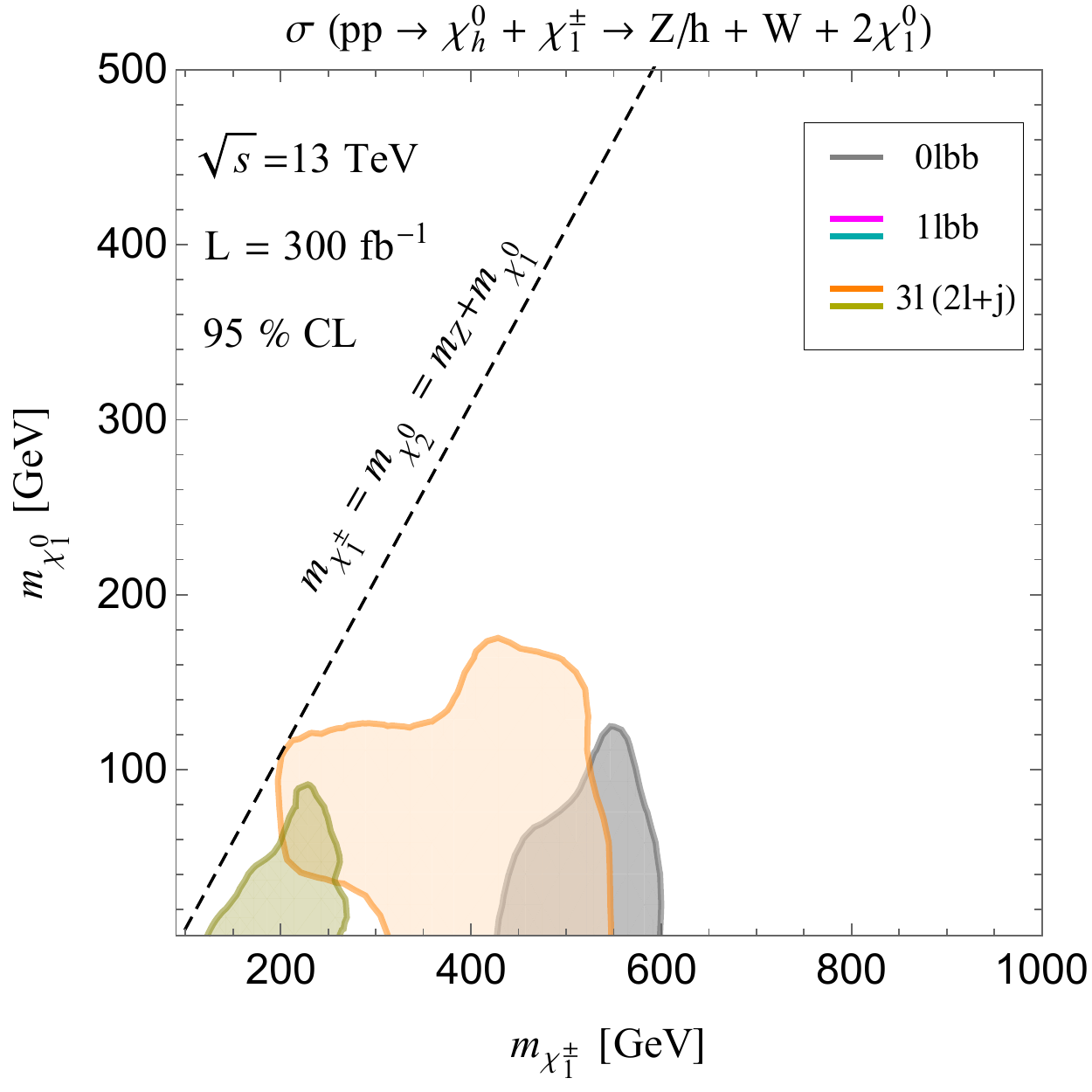} 
	\includegraphics[width=0.48 \columnwidth]{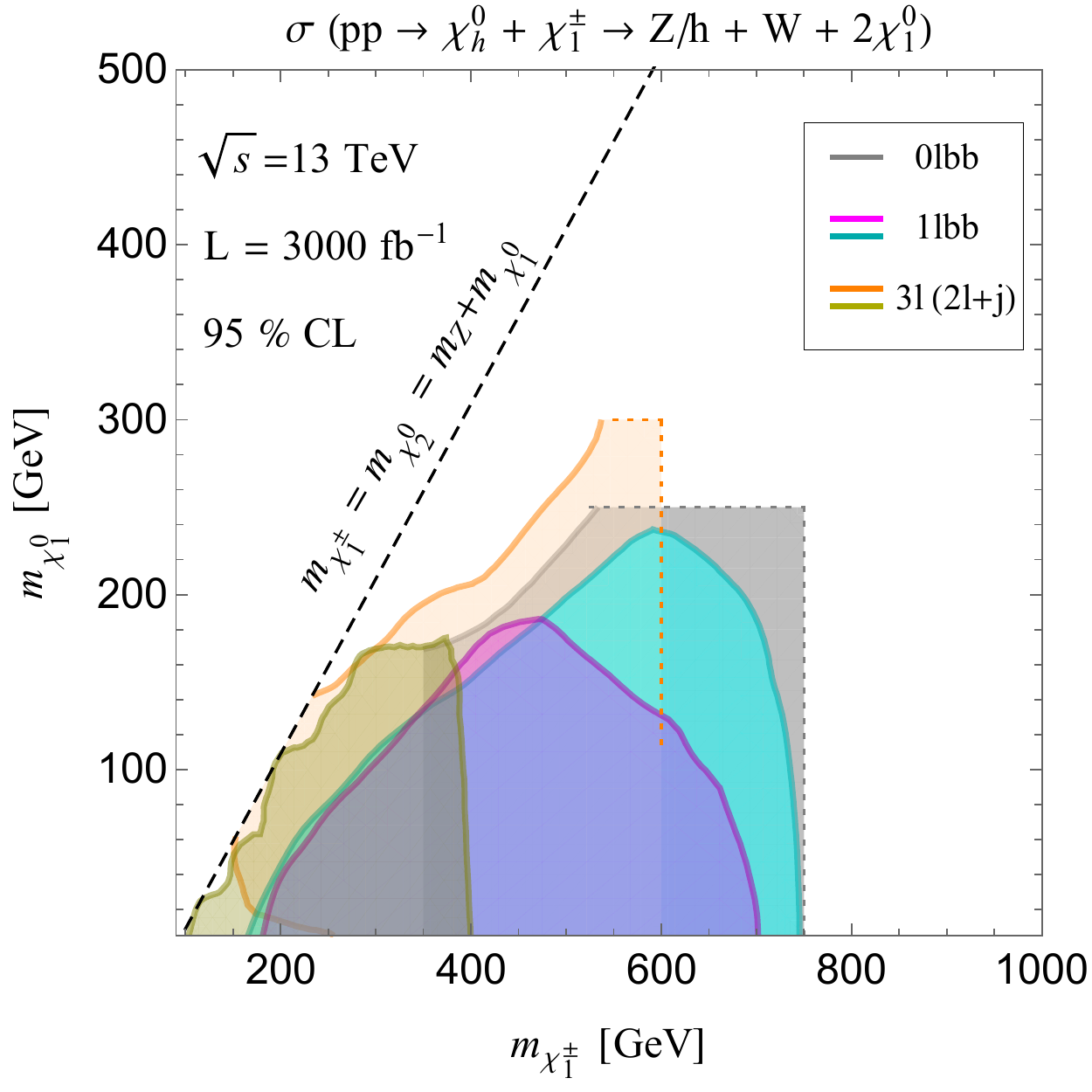} \\
	\includegraphics[width=0.48 \columnwidth]{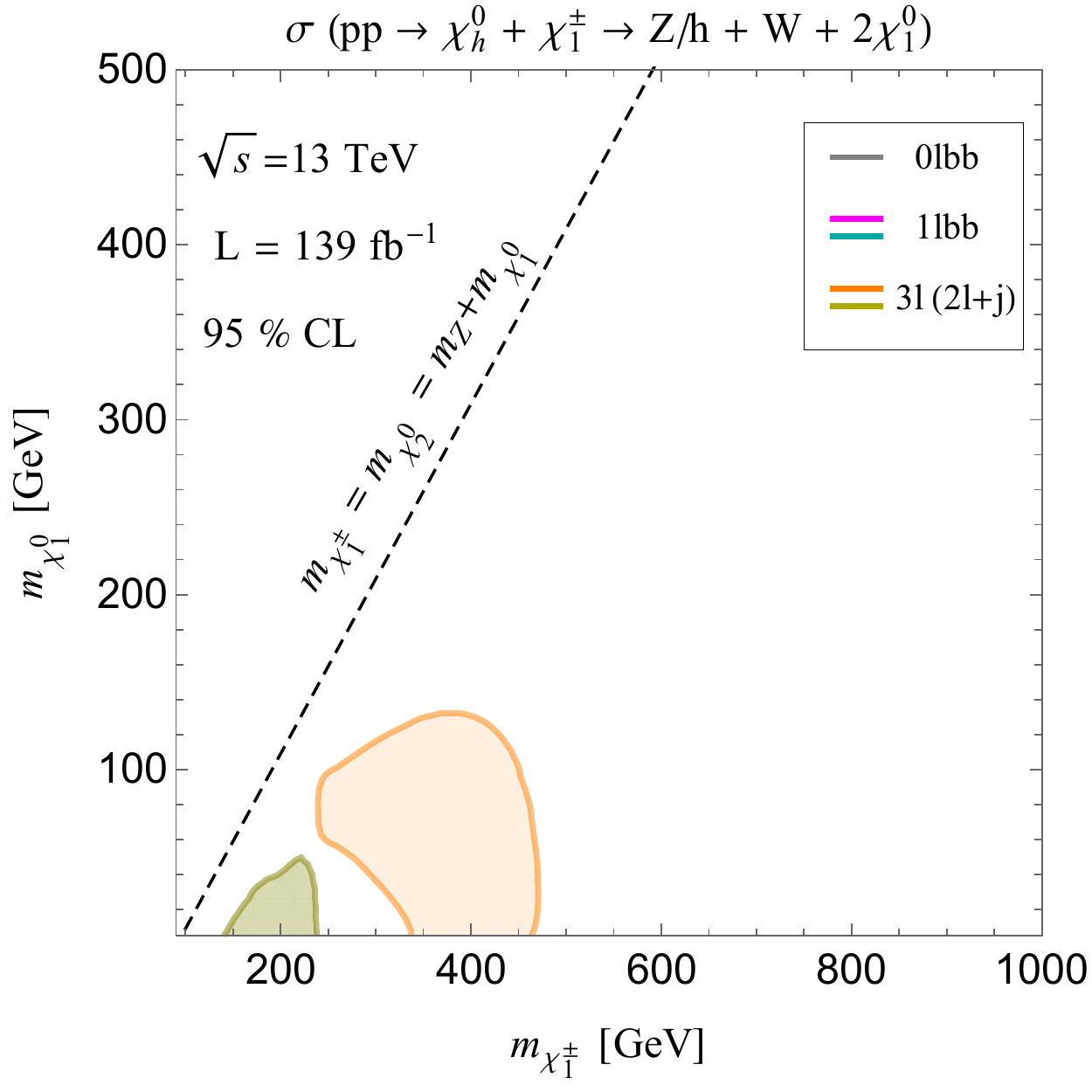} 
	\includegraphics[width=0.48 \columnwidth]{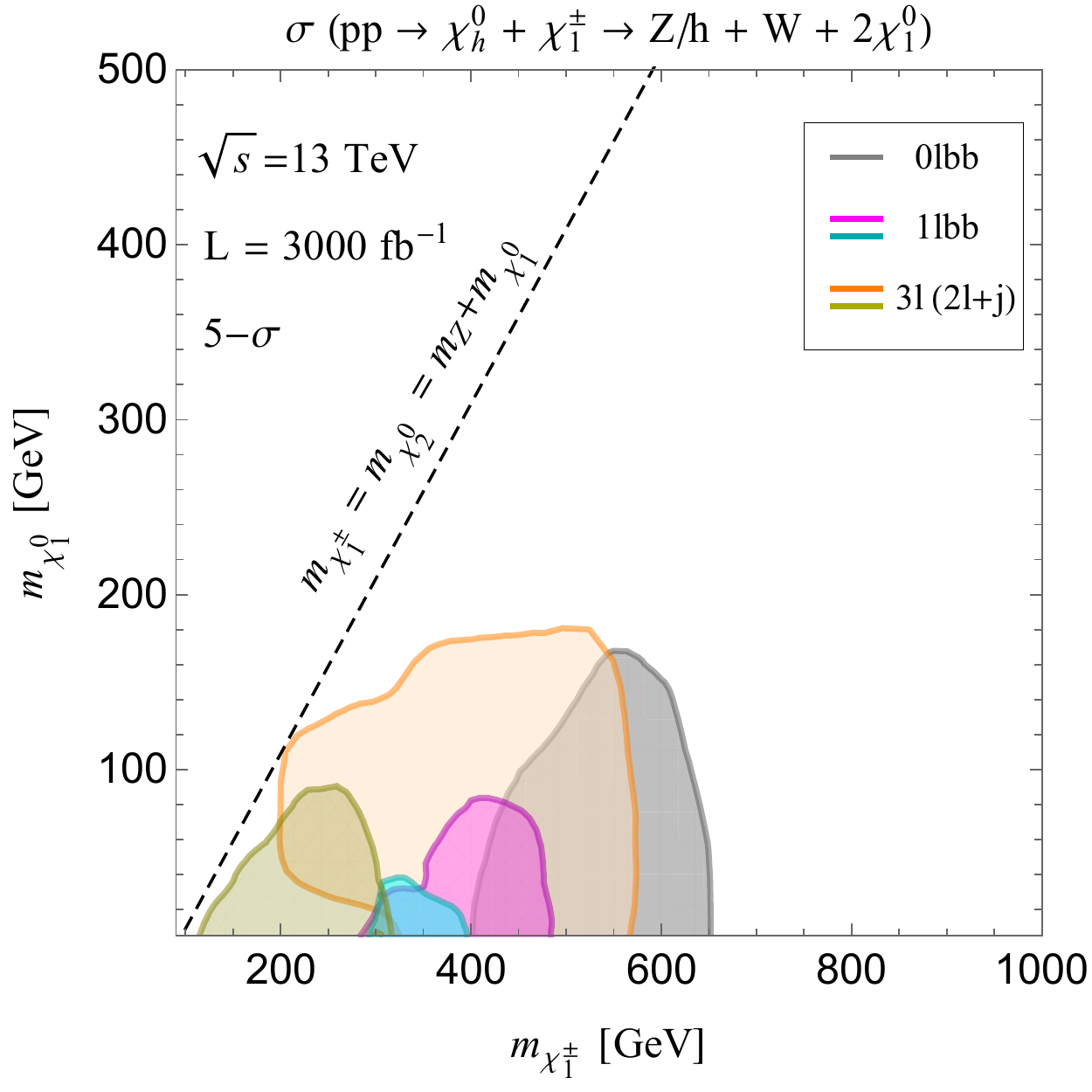} 
\caption{Constraints on the Bino-Higgsino scenario projected to 300 and 3000 fb$^{-1}$ at $95\%$ confidence level in the top panel, 139 fb$^{-1}$ at $95\%$ confidence level and 3000 fb$^{-1}$ at $5~\sigma$ in the lower panel. We choose $\tan\beta = 5$, but the results depend very weakly on this choice for the range of parameters that we explore. The $0\ell bb$ (gray)~\cite{Aaboud:2018ngk} and $1\ell bb$ (magenta, cyan)~\cite{Aad:2019vvf} come from the $\chi_{2,3}^0 \chi_1^\pm\to hW+2\chi_1^0$ channel, with $h\to \bar{b}b$ and $W$ decay to hadronic or leptonic final states. The $3\ell$ (dark yellow)~\cite{ATLAS-CONF-2019-020}  and $3\ell/2\ell+\text{j}$ (orange)~\cite{ Aaboud:2018sua}  come from the $\chi_{2,3}^0 \chi_1^\pm\to ZW+2\chi_1^0$ channel, with $Z \to 2\ell$. For the gray and orange shaded region, there are dotted lines cutting off the high mass region because $\sigma_{\rm limit}$ is not provided from HEP-Data. }
\label{fig:constraints}
\end{figure} 

This is the traditional search channel for electroweakinos and leads to very powerful results at
high luminosities. The ATLAS collaboration has explored this channel using a variety of techniques and final states. As mentioned, final states with three leptons \cite{ATLAS-CONF-2019-020} or two leptons with jets \cite{ Aaboud:2018jiw} provide the strongest reach. After recasting the LHC analyses for the case of Higgsinos and using both
the Recursive Jigsaw reconstruction method as well as the standard ones, we observe the sensitivity for Higgsino-like electroweakinos
at a luminosity of 300~fb$^{-1}$ in the top left panel of Fig.~\ref{fig:constraints}. The LHC should be capable of reaching Higgsino
masses of about 500~GeV, but only for lightest neutralino masses that are lower 
than 200~GeV, and far from the kinematic threshold for the decay of the second
and third lightest neutralino into $Z$ states.  This is due to the fact that, for masses close
to this kinematic threshold, the leptons carry low $p_T$ values.  This is one of
the main advantages of the Higgs production mode, where the leptons will receive higher
$p_T$ contributions due to the high $p_T$ of the electroweakinos proceeding from the
decay of the heavy Higgs bosons.

In the top right panel of Fig.~\ref{fig:constraints}, at higher luminosity of 3000~fb$^{-1}$, the situation is highly improved.
On one hand, Higgsino masses up to 750~GeV could be reached in this case, for
lightest neutralino masses up to about 300~GeV. The reach for $3\ell$ (dark yellow line) stops at
the dotted lines. In this case, $\sigma_{\rm limit}^{\rm up}$ is not provided for the higher mass region in the current searches. However, it is expected that updated searches would be sensitive to this region.
In general, this is a highly interesting reach,
that covers the region of Higgsino masses one would associate with a natural low
energy supersymmetry scenario. But the kinematic limitations close to threshold
are still present, although only for higher lightest neutralino masses.

\subsection{hW Channels}

A secondary channel to probe the production of charginos and neutralinos occurs when the neutralinos decay to a SM Higgs boson. The ATLAS collaboration has studied this channel for electroweakino pair production of $\chi_1^\pm \chi_2^0$. Similarly, the production cross sections assumed couplings associated to Wino-like couplings and the decay branching ratio of $\chi_1^\pm  \to W^\pm \chi_1^0$ and $\chi_2^0  \to h \chi_1^0$ are both $100\%$ \cite{Aad:2019vvf, Aaboud:2018ngk}. The final states considered in these studies are $0\ell + \bar{b}b $, $1\ell + \bar{b}b $,
$1\ell + \gamma\gamma $ and $\ell^\pm \ell^\pm $, which arise from a combination of leptonic and hadronic decays of the $W^{\pm}$, and SM Higgs decay into $\bar{b}b$, $\gamma\gamma$ and $2\ell$. From their results we find that the $0\ell + \bar{b}b $ and $1\ell + \bar{b}b$ channels provide the strongest bound, therefore to explore the ultimate reach for Higgsino-like electroweakinos we only recast these channels.

We find that these channels provide a $95\%$ CL reach for larger chargino masses almost up to 600 (750) GeV for luminosity of 300 (3000)~fb$^{-1}$, as shown in the top panel of Fig.~\ref{fig:constraints}. 
For luminosity of 300~fb$^{-1}$, only the $0\ell + \bar{b}b $ channel provides
constraints, for $m_{\chi_1^\pm} $ between 400--600 GeV and very light $\chi_1^0$, while the bound from the $1\ell + \bar{b}b $ channel vanishes. This already shows the need for higher luminosities to search for higgsino-like electroweakinos 
in this decay channel.  For luminosity of 3000~fb$^{-1}$, we see a dramatic enhancement of the constrained region for both $0\ell + \bar{b}b $ and 
$1\ell + \bar{b}b $ channels. 

Moreover, in the bottom right panel of Fig.~\ref{fig:constraints} we show the $5\sigma$ reach for direct production of Higgsino-like states at HL-LHC both for the $ZW$ and $hW$ channels. Comparing with the bottom left panel of Fig.~\ref{fig:constraints},  we see that it is still possible to discover light electroweakinos 
in the HL-LHC era in regions of parameters not already probed at current collected luminosity of 139~fb$^{-1}$. Our results are consistent with other projections to future upgrades at the LHC~\cite{ Han:2013kza,ATLAS:2013hta, CMS:2013xfa}. \footnote{For other theoretical studies in this direction see \cite{Han:2016qtc}.
}

\section{Reach of the Higgs decay production mode at higher luminosities}
\label{sec:search-from-heavy-Higgs}

Although there are already a few theoretical analyses of this mode~\cite{Gori:2018pmk}--\cite{Baum:2019uzg}, 
there is to our knowledge no
dedicated search for electroweakinos in this production mode.  Therefore we performed our 
own analysis of the background and production rates \footnote{We have closely followed the analysis strategy given in Ref.~\cite{Gori:2018pmk}, where the same signature has been studied exclusively in the compressed region, $m_{\chi_{2,3}^{0}} - m_{\chi_{1}^{0}}\simeq m_{Z}$ and we find a similar reach.}.
We concentrated only on the gluon fusion mode, and on the lepton decays of the $Z$ bosons
proceeding from the decay of the second and lightest neutralinos in the lepton plus missing
energy analysis. Thus, the relevant signal results in two opposite-sign, same flavor leptons with significant missing energy

\begin{equation}
gg\rightarrow H/A \rightarrow \chi_{h}^{0} + \chi_{1}^{0}\rightarrow \ell^{+} \ell^{-} + \slashed{E}_{T},
\end{equation}
where we sum the channels from heavy neutral scalar $H$ and pseudoscalar $A$ production, and $h=2,3$. 

The production of Higgs bosons become significant at either small or large values of $\tan\beta$,
for which the gluon fusion cross section is proportional to the large top and bottom-quark 
Yukawa couplings, respectively.  The main couplings governing the production cross section
are given by
\begin{equation}
g_{H t \bar{t} } \simeq  g_{A t \bar{t}} \simeq \frac{m_t}{v \tan\beta}
\end{equation}
and 
\begin{equation}
g_{H b \bar{b}} \simeq g_{A b \bar{b}} \simeq \frac{m_b \tan\beta}{v (1 + \Delta_b)},
\end{equation}
where $m_t$, $m_b$ and $v$ are the running top-quark and bottom-quark masses and
the Higgs vacuum expectation value $v \simeq 174$~GeV, respectively. The factor
$\Delta_b$ is associated with a threshold correction~\cite{Carena:1998gk,Carena:2016npr}, induced by the decoupling
of the supersymmetric particles~\cite{Hempfling:1993kv,Hall:1993gn,Carena:1994bv}, which tends to be suppressed for low values of
the Higgsino mass parameter compared to the heavy colored particles. Moreover,
$\Delta_b$ becomes significant in this case only for very large values of $\tan\beta$,
for which the heavy Higgs bosons are heavily constrained via their decay
into $\tau$ leptons~\cite{Sirunyan:2018zut,Aad:2020zxo}.

Regarding branching ratios, it is important to remark that the same couplings that
control the production cross sections also control the branching ratios into top and
bottom quarks. Hence, assuming Higgs masses above the top quark pair
production threshold, the gain in production
cross section at low and large values of $\tan\beta$ is compensated
by the reduction of the branching ratio of the decay into electroweakinos.
Overall, considering both the Higgs production and decay branching ratios
while concentrating on the gluon fusion analysis, we find that the total 
production rate coming from the decay into electroweakinos is 
enhanced for lower values of $\tan\beta$ and  decreases slowly for larger
values of $\tan\beta$.

In the MSSM, for moderate values of the Higgsino mass parameter $\mu$, low values of $m_A$ lead 
to a modification of the Higgs couplings which therefore affect the agreement between the theoretical
predictions and the Higgs precision measurements~\cite{Carena:2014nza,Haber:2017erd}. The modifications of the 
Higgs couplings are approximately independent of $\tan\beta$ and lead to a bound of about 600~GeV
on $m_A$~\cite{Bahl:2018zmf}. Observe, however, that these bounds may  be modified considering
the low energy theory to proceed from, for instance, the NMSSM in the limit of heavy singlet~\cite{Carena:2015moc},  
that may arise dynamically from the renormalization group evolution at low energies~\cite{Benakli:2018vqz,Coyle:2019exn}.
In this article,  we have mainly focused on the region of masses higher than 600~GeV, since they 
allow to avoid the precision measurement constraints in the MSSM, but also enhance the reach for
heavier electroweakinos.  

Beyond the precision measurement constraints, other relevant constraints come from the search for heavy Higgs bosons 
decaying into SM final states. 
As discussed before, in the MSSM, the decays of the heavy Higgs to bottom quarks and tau leptons are enhanced by large  $\tan\beta$. The $H/A \to \tau^+ \tau^-$ decay channel has been studied in the ATLAS Run-II search ($139 ~\text{fb}^{-1}$) \cite{Aad:2020zxo} in the case where at least one tau lepton decays hadronically and assuming that all 
superpartners are heavy enough that the $H/A$ Higgs bosons can not decay into them. In this case, the current bounds would exclude values of $\tan\beta\gtrsim 8 $ for $m_{A} = 600$ GeV. However, in the present case the heavy neutral Higgses can decay into a Higgsino-Bino pair with branching ratios close to $30\%$ which, for given $m_{A}$, alleviates the upper limit on $\tan\beta$.

Recently, the comparison of the current and projected bounds between these two scenarios has been studied in \cite{Bahl:2020kwe}. They find that for scenarios with light electroweakinos the bound from $H/A \to \tau^+ \tau^-$ excludes $\tan\beta \gtrsim 15$, for $m_{A} = 600$ GeV and luminosity of  $36~\text{fb}^{-1}$. While for the HL-LHC at 14 TeV and luminosity of $3~\text{ab}^{-1}$, they project this bound to be $\tan\beta \gtrsim 8$. Additionally, constraints for the $H/A \to \bar{b}b$ decay have been presented in \cite{Aad:2019zwb}. However, the bound is quite weak and only excludes values of $\tan \beta > 20$, for the same region of heavy Higgs masses. As we shall see, irrespectively of these bounds, the sensitivity of the 
search for electroweakinos proceeding from the decay of heavy Higgs bosons is highly reduced for values of $\tan\beta > 10$, even for low masses of the heavy Higgs.

For lower $\tan \beta$, there exist constraints from the decays of the charged Higgs bosons, $H^+ \to t\bar{b}$, and also the neutral Higgs channel, $H/A \to \bar{t}t$. In the case of the charged Higgs the bounds become weak in the region of parameters we are considering, as can be seen from the results of Ref.~\cite{Aad:2020kep}. For instance, for $\tan\beta = 0.5$ the mass of the charged Higgs should satisfy $m_{H^\pm} > 1.2$ TeV, while for $\tan\beta = 1$ there are no constraints for $m_{H^\pm}$. 
For the decay of the neutral Higgs bosons to $\bar{t}t$, the projected HL-LHC
sensitivity is $\tan\beta \lesssim 2$ for a heavy Higgs mass $m_A = 600$~GeV~\cite{Carena:2016npr}. However, considering top associated production can mildly increase this sensitivity to $\tan\beta \lesssim 3$. 
The current limits for the $\bar{t}t$ channel for $m_A=600$~GeV is about $\tan\beta \lesssim 1.5$, coming from the analysis of the CMS 13 TeV $36 ~{\rm fb}^{-1}$ data~\cite{Sirunyan:2019wph}. 
In our analysis, we shall limit our analysis to values of $\tan\beta \geq 2$.  Also, unless the stop masses are larger
than $10^5$~GeV, it is impossible to obtain the proper Higgs mass in the MSSM for values of $\tan\beta$ lower than 2~\cite{Draper:2013oza,Bagnaschi:2014rsa,Vega:2015fna,Lee:2015uza,Bahl:2017aev}.

Another possible search channel for the heavy Higgs follows from the same production mechanism but followed by the heavier neutralinos decaying instead to the SM Higgs boson, $\chi_{2,3}^{0}\rightarrow h + \chi_{1}^{0}$, followed by the Higgs decay to bottom quarks. Despite the fact that this decay chain can have a sizable rate, depending on $m_{A}$ and $\tan\beta$, this signal suffers from multiple QCD backgrounds. We have estimated the signal significance of this channel using the background and analysis of Ref. \cite{ATLAS:2018bvd}. We find that the reach of this channel will not be competitive in comparison to direct searches or the dilepton Higgs channel. Thus, we will not explore this possibility in this work.

\begin{figure}[htb]
	\centering
	\includegraphics[width=0.45 \columnwidth]{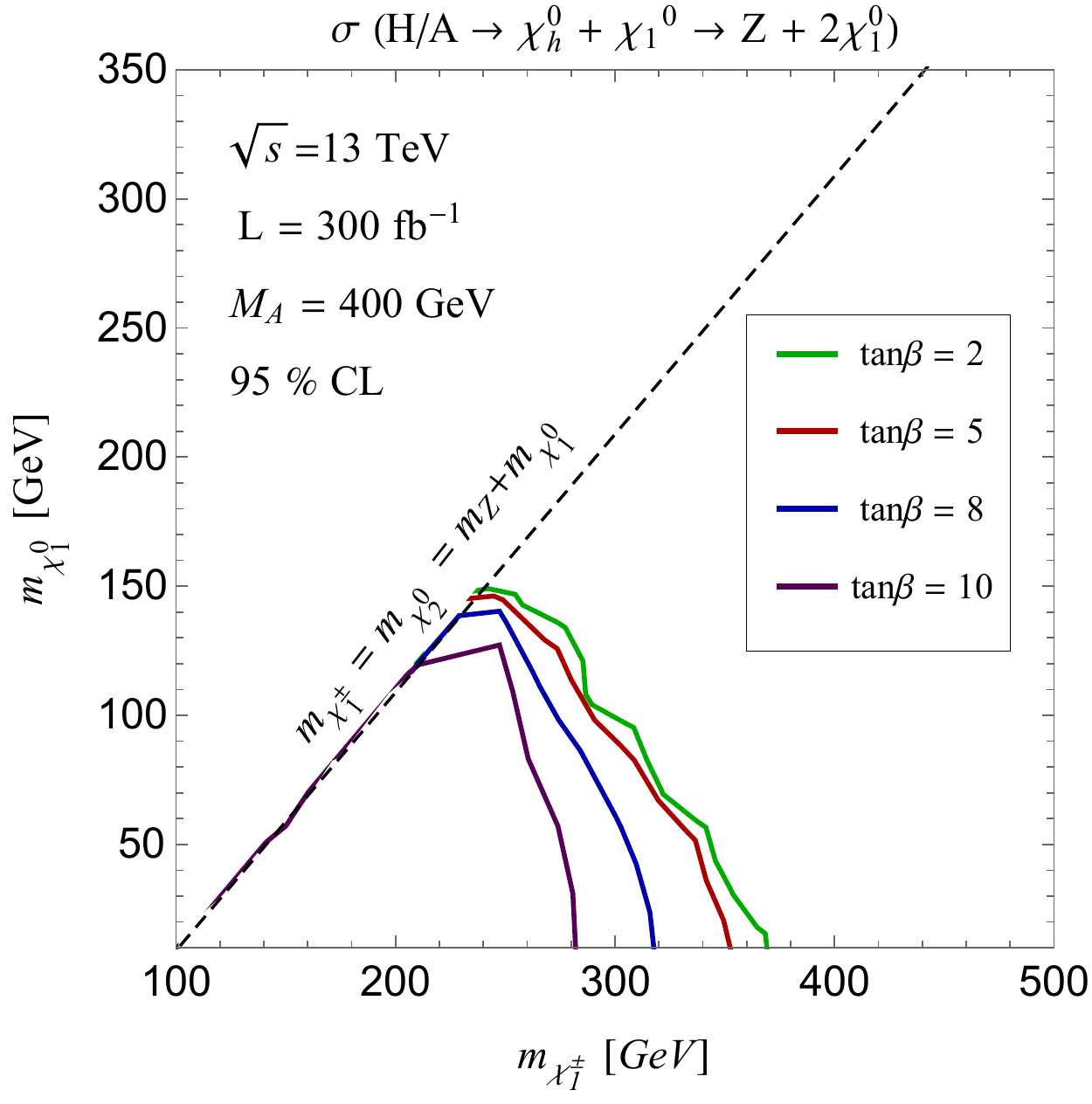}
	\includegraphics[width=0.45 \columnwidth]{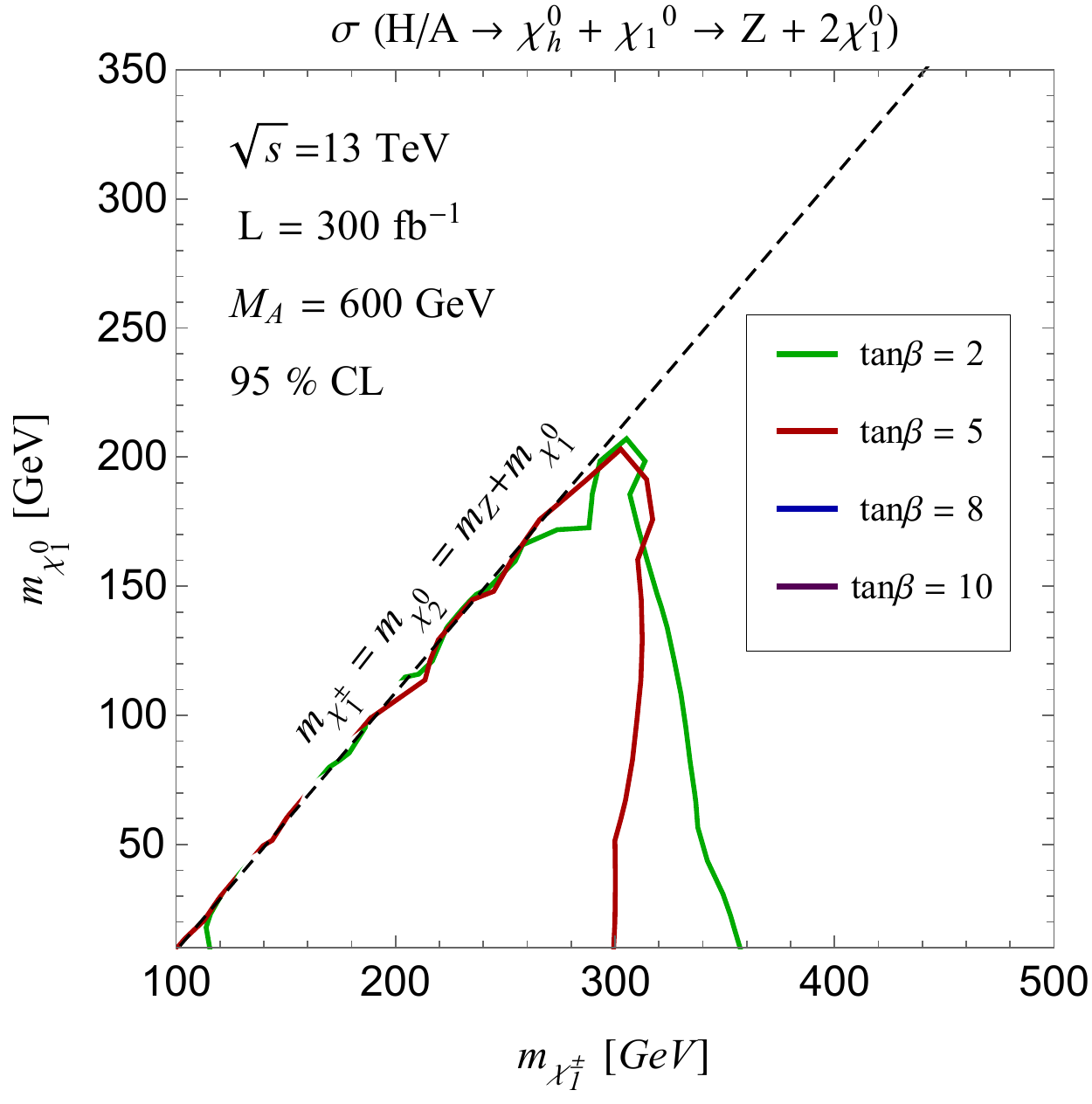}
	\caption{95 \% CL bounds on the  heavy neutral Higgs decay into electroweakinos $H/A \to \chi_h^0 \chi_1^0 \to Z+2\chi_1^0$, with $Z \to \ell^+ \ell^-$. With integrated luminosity of $300~{\rm fb}^{-1}$ and $m_{H/A} = 400~\&~600 $ GeV, the sensitivities for electroweakinos are shown as contours in the $m_{\chi_1^{\pm}}$-- $m_{\chi_1^0}$ plane for $\tan\beta = 2, 5, 8,~\&~ 10$. For $m_{H/A} = 600 $ GeV the bounds vanish for values of $\tan\beta \gtrsim 5$.}
	\label{fig:higgs_bounds_300fb}
\end{figure}

\begin{figure}[h!]
	\centering
	\includegraphics[width=0.48 \columnwidth]{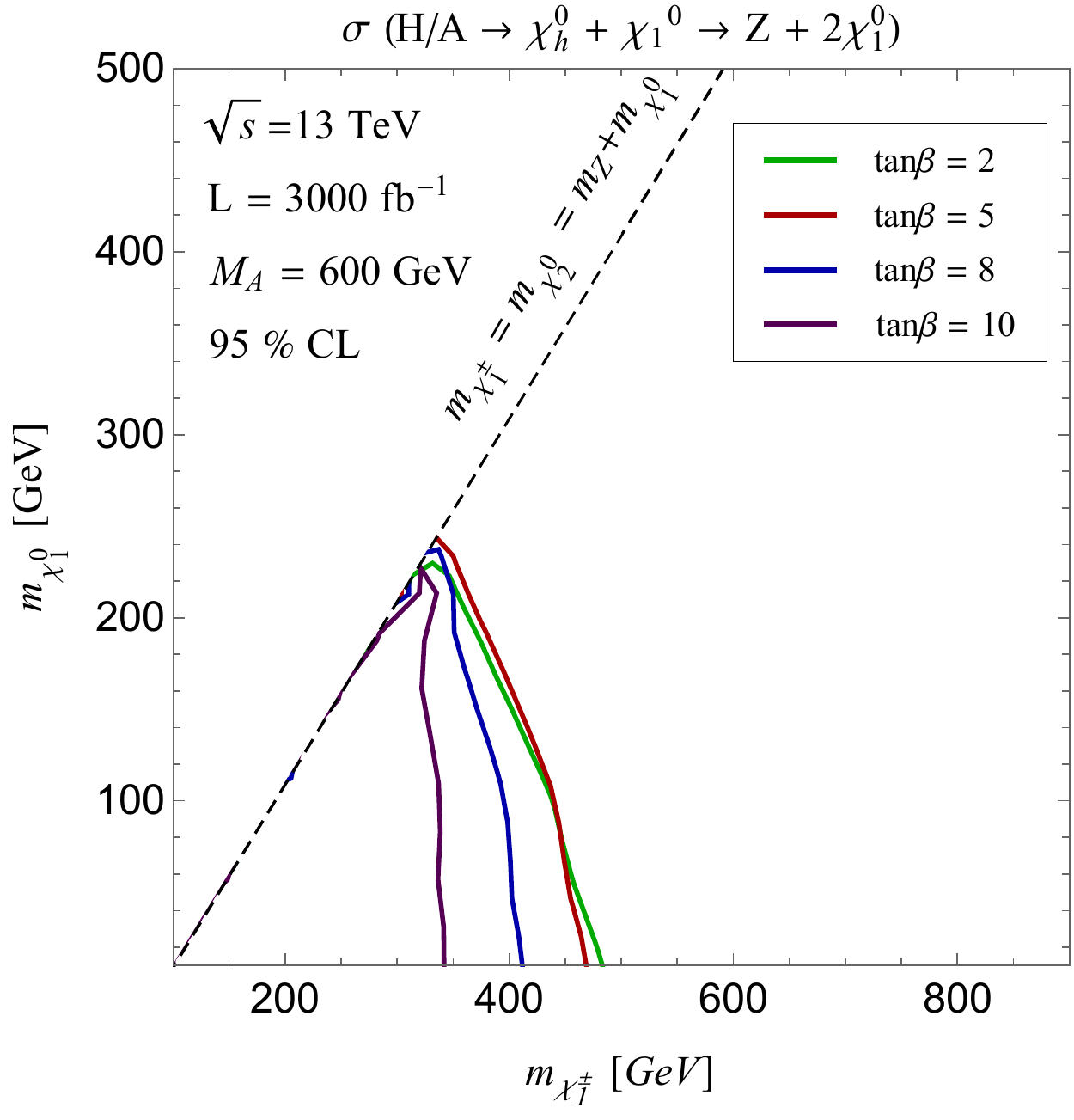}
	\includegraphics[width=0.48 \columnwidth]{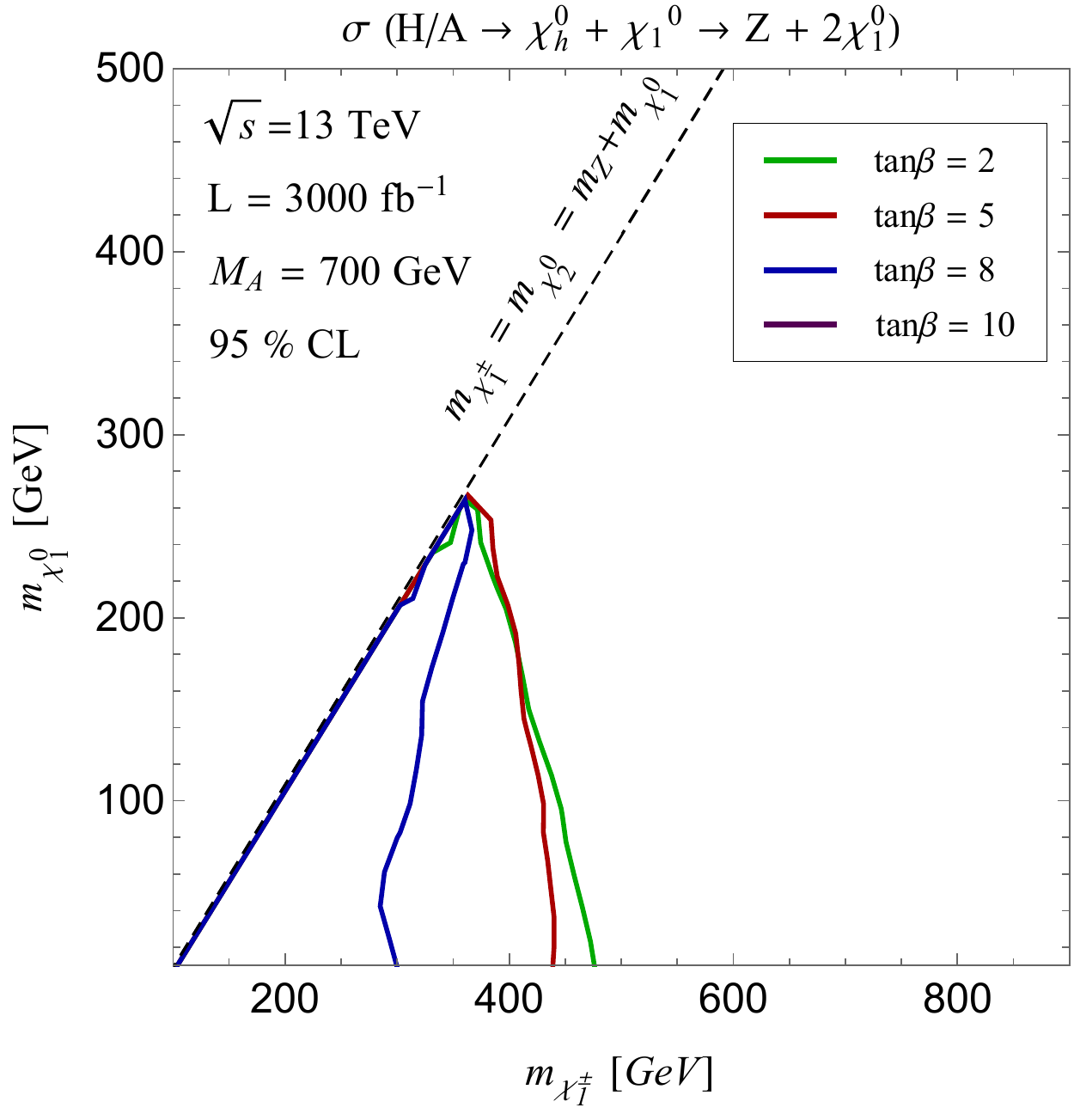} \\
	\includegraphics[width=0.48 \columnwidth]{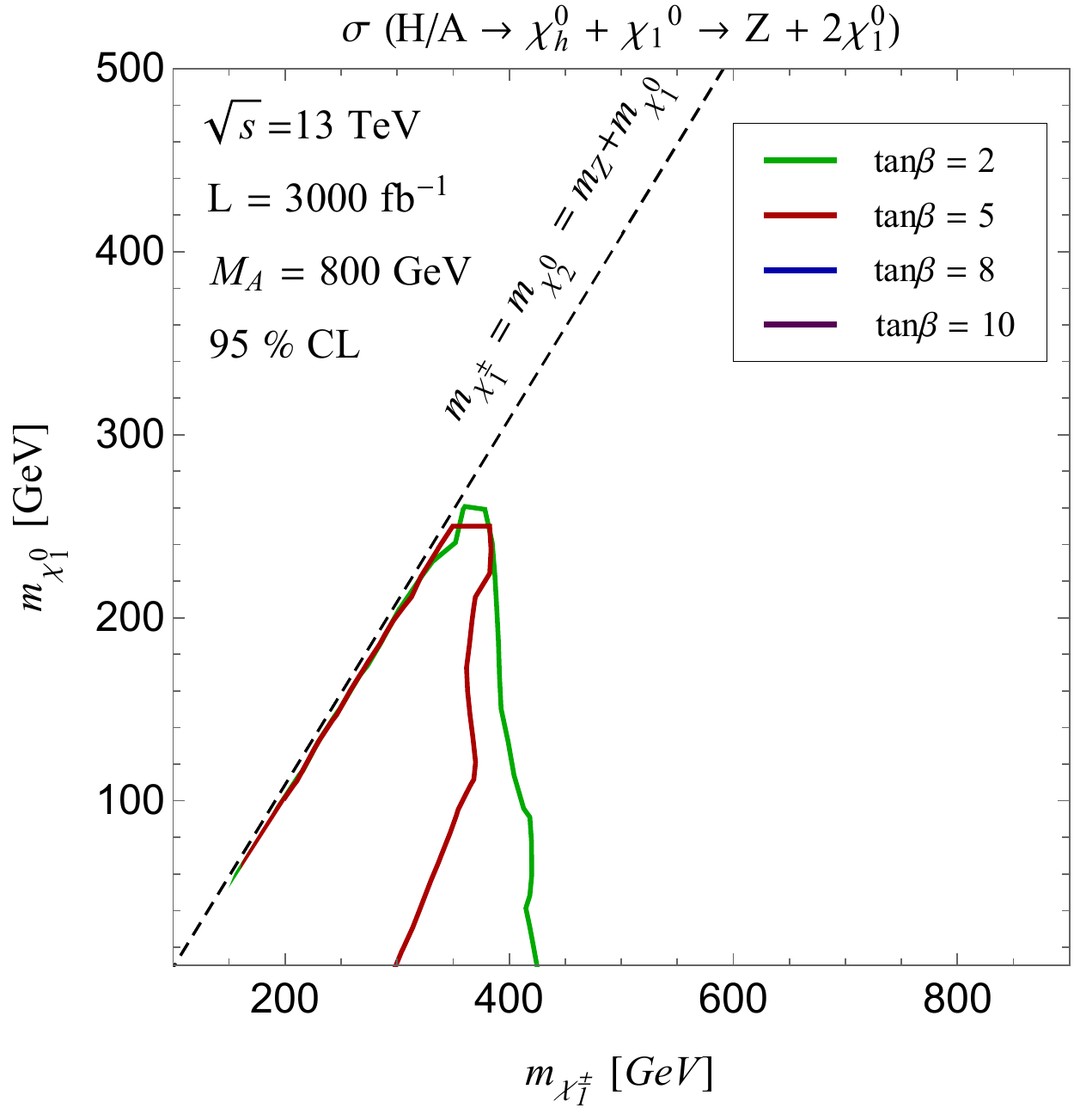}
	\includegraphics[width=0.48 \columnwidth]{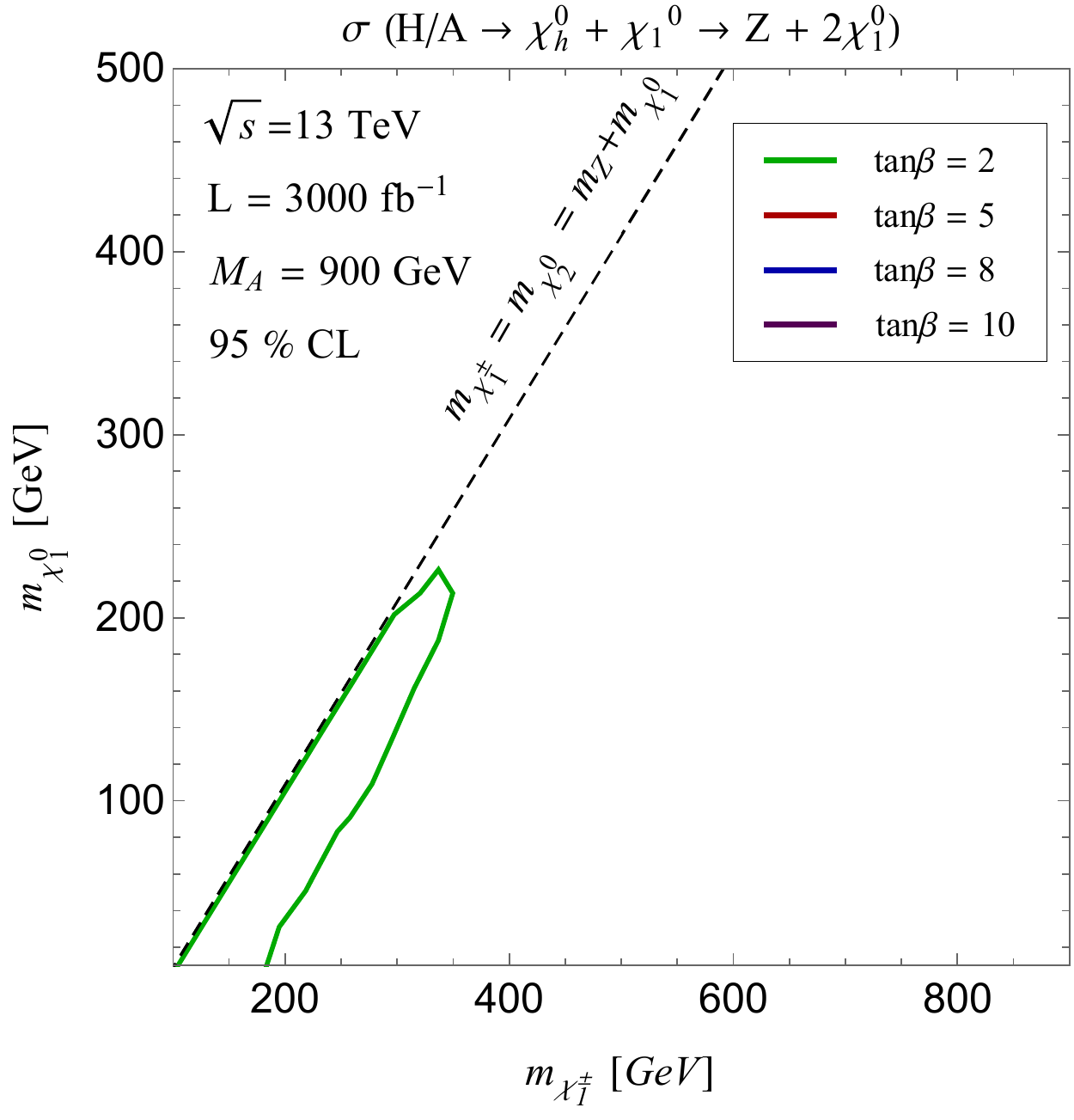} 
	\caption{95 \% CL bounds on the  heavy neutral Higgs decay into electroweakinos $H/A \to \chi_h^0 \chi_1^0 \to Z+2\chi_1^0$, with $Z \to \ell^+ \ell^-$. With integrated luminosity of $3000~{\rm fb}^{-1}$ and $m_{H/A} = 600,~ 700,~ 800,~900 $ GeV, the sensitivities for electroweakinos are shown as contours in the $m_{\chi_1^{\pm}}$--$m_{\chi_1^0}$ plane for $\tan\beta= 2-10$.}
	\label{fig:higgs_bounds_3000fb}
\end{figure} 

To compute the resulting bounds, requiring that events contain a significant amount of missing energy drastically reduces the contribution from QCD backgrounds. The relevant backgrounds then come from di-boson production, $pp\rightarrow VV$, with $V=W,Z$. Further requiring that one pair of opposite-sign, same flavor leptons reconstruct the $Z$-boson mass, $85\text{ GeV}< m_{\ell\ell} < 95\text{ GeV}$, leaves the $ZZ$ channel as the most relevant source of background. Apart from large $\slashed{E}_T$ in the events, the signal significance can be further improved by using the modified clustered transverse mass introduced in Ref.~\cite{Gori:2018pmk},

\begin{equation}
m_{cT}^{2}(\ell\ell,\slashed{E}_{T})=2\times \left((|\vec{p}_{T}^{~\ell\ell}| + |\vec{\slashed{p}}_{T}|)^{2} - |\vec{p}_{T}^{~\ell\ell} + \vec{\slashed{p}}_{T}|^{2}\right),
\end{equation}
where $\vec{p}_{T}^{~\ell\ell}$ is the transverse momentum of the di-lepton system, and $\vec{\slashed{p}}_{T}$ is the two-vector of the missing transverse energy. For more details of the kinematics and impact of the cuts see Ref.~\cite{Gori:2018pmk}. We generate background and signal using {\tt MadGraph5} followed by shower and detector simulation with {\tt Pythia8} \cite{Sjostrand:2006za, Sjostrand:2014zea} and {\tt Delphes} \cite{deFavereau:2013fsa} respectively. For a given heavy Higgs mass and $\tan\beta$, we calculate the signal significance in the chargino-neutralino plane, calculated from Eq.~(\ref{eq:params}), by optimizing lower and upper cuts on $\slashed{E}_T$ and $m_{cT}^{2}$, respectively.

Our results for 300~fb$^{-1}$ are presented   in Fig.~\ref{fig:higgs_bounds_300fb} in the 
$m_{\chi_1^{\pm}}$--$m_{\chi_1^0}$ plane.  Viewing the bounds   in this way allows for a straightforward comparison with the
direct production search channels in the same parameter space.  For 300~fb$^{-1}$,
the LHC sensitivity for heavy Higgs bosons decaying into electroweakinos is limited to relatively low values of $m_A$, that as argued before 
would lead to tension with Higgs precision measurements, unless one goes beyond the pure MSSM description. 
In the left panel, we show the bounds for $m_{A}=$~400~GeV. For $\tan\beta = 2$, the sensitivity covers almost the entire region that is kinematically allowed. Whereas for $\tan\beta = 10$, the sensitivity reaches to chargino masses of almost 300~GeV, and slightly above 100~GeV for the lightest neutralino mass. In the right panel, in the case of $m_{A}=600$ GeV, we see that 
the sensitivity is limited to low values of $\tan\beta = 2$ and 5. For larger values of the Higgs masses the bounds become even weaker and vanish for $m_A \gtrsim 700$~GeV, due to the decrease in the  production cross-section
$\sigma( gg \to H/A \to \chi_h^0 \chi_1^0 \to Z+2\chi_1^0)$. 

In Fig.~\ref{fig:higgs_bounds_3000fb}, we show the reach for 3000~fb$^{-1}$.
We find that for heavy Higgs masses of 600--700GeV, the reach can explore a broad range of values of $\tan\beta$. Further, for larger masses of the heavy Higgses the reach of the Higgs decay channel generally improves in the region of the kinematic limit of the $\chi_{2,3}^0 \to Z+\chi_1^0$ decay. This is expected as the cut efficiency of the kinematics resulting from high $p_T$ leptons compensates the decrease in cross sections.
For Higgs masses above $1$ TeV, we find no reach for any values of $\tan\beta$, since beyond this point the efficiency gain cannot compete with the drop in the production cross section.

\begin{figure}[htb]
	\centering
	\includegraphics[width=0.48 \columnwidth]{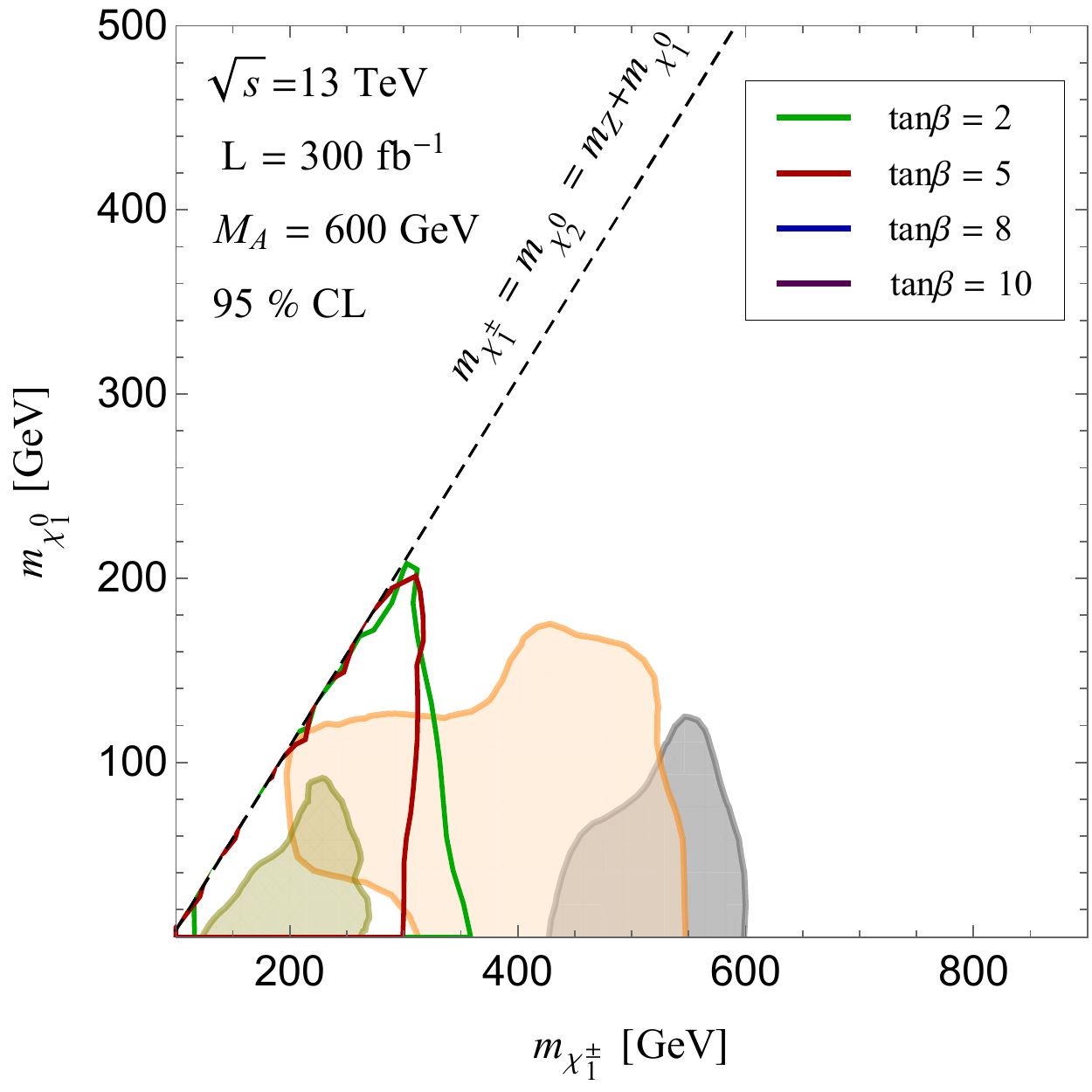}
	\includegraphics[width=0.48 \columnwidth]{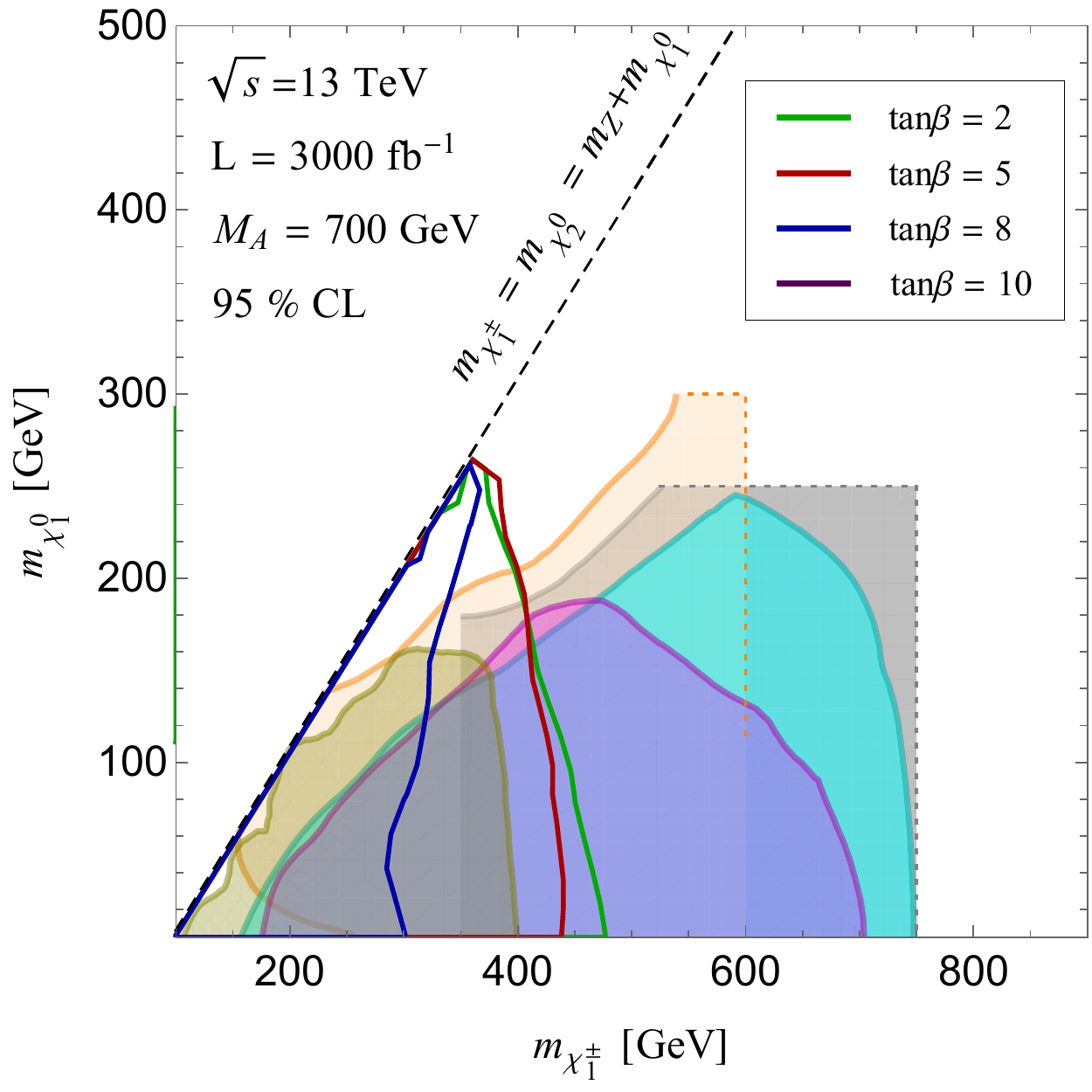}
	\caption{Comparison between the $H/A \to \chi_h^0 \chi_1^0 \to Z+2\chi_1^0$ electroweakino search and the existing direct production constraints for the higgsino-like electroweakinos at HL-LHC.  In the case of the resonant Higgs channel we present the bounds for $m_A = 600 ~(700)$ GeV for luminosity of $300 ~(3000)\text{ fb}^{-1}$.}
	\label{fig:higgs_bounds-compare}
\end{figure} 

In Fig.~\ref{fig:higgs_bounds-compare}, we compare the $H/A \to \chi_h^0 \chi_1^0 \to Z+2\chi_1^0$ search to existing direct production constraints interpreted for higgsino-like electroweakinos,
as presented in the top right panel of Fig.~\ref{fig:constraints}. The sensitivities are nicely complementary to each other. We find that there is a significant overlap in the parameter space 
that can be covered between the proposed Higgs decay search and the direct production searches. We note that since there is still a
chance to obtain a $5 ~\sigma$ discovery at HL-LHC, as we show in the bottom right panel of Fig.~\ref{fig:constraints}, the search for electroweakinos will become an alternative way of probing
the existence of a heavy Higgs boson beyond the traditional decay channels to $\tau$-leptons, and  top and bottom quarks. As emphasized  before,
the proposed Higgs decay channel also has better sensitivity close to the compressed mass region $m_{\chi_1^{\pm}, \chi_h^0} \simeq m_{\chi_1^0} + Z$ compared with the direct production case, offering a unique channel to explore this region of electroweakino masses.

\section{Conclusions}
\label{sec:conclusion}

In this article we have analyzed the search for electroweakinos at the LHC, putting emphasis on the complementarity
of direct and Higgs decay production modes. We have considered the case of heavy scalar superpartners and 
concentrated on the well motivated case of light Higgsinos and Binos.   We have shown that the LHC reach in this case
remains weak in both production channels at a luminosity of 300~fb$^{-1}$ but becomes very promising  at higher luminosities.

In the direct production mode, the high luminosity LHC, with a total integrated luminosity of 3000~fb$^{-1}$ will be able to cover 
up to masses of the lightest 
chargino and second and third lightest neutralino states up to the order of 800 GeV for lightest neutralinos lower than about 
300 GeV. This is comparable to the current reach at the LHC for the case of light Winos and significantly larger than the
reach for light Higgsinos at 300~fb$^{-1}$, that go up to chargino masses of about 600~GeV for lightest neutralinos not
heavier than 150~GeV.  In spite of the considerable reach, a large gap is open when the mass differences between
the heavier electroweakino states and the lightest neutralino state become small. 

The reach in the heavy Higgs decay mode depends strongly on the Higgs boson masses, and cannot go up to chargino
masses as large as in the direct production mode. However, for heavy Higgs boson masses larger than 500 GeV
and smaller than about 900~GeV, and moderate values of $\tan\beta$, it can go up to chargino masses of about 500~GeV 
for lightest neutralino masses not heavier than 300 GeV. Interestingly, although the coverage is not as strong as the
direct production case, this search is not limited by the same kinematic considerations as the ones in the direct production
mode and can cover masses up to the kinematic threshold for the decay of the heavier electroweakinos into the lightest 
neutralino.  Moreover, this mode provides also an alternative way of looking for heavy Higgs bosons in this range
of masses and hence should be a high priority for future LHC analyses.

\section*{Acknowledgments}
We would like to thank S. Gori and N. Shah for very useful discussions and communication. 
Work at University of Chicago is supported in part by U.S. Department of Energy grant number DE-FG02-13ER41958. Work at ANL is supported in part by the U.S. Department of Energy under Contract No. DE-AC02-06CH11357. JL acknowledges support by Oehme Fellowship. During the course of this work NM has been supported by the U.S. Department of Energy, Office of Science, Office of Work- force Development for Teachers and Scientists, Office of Science Graduate Student Research (SCGSR) program. The SCGSR program is administered by the Oak Ridge Institute for Science and Education (ORISE) for the DOE. ORISE is managed by ORAU under contract number de-sc0014664.

\appendix
\section{Auxiliary figures}

\begin{figure}[htb]
\centering
	\includegraphics[scale=0.5]{figures/MA_600_higgs_xsection_mu_minus.pdf} 	\includegraphics[scale=0.5]{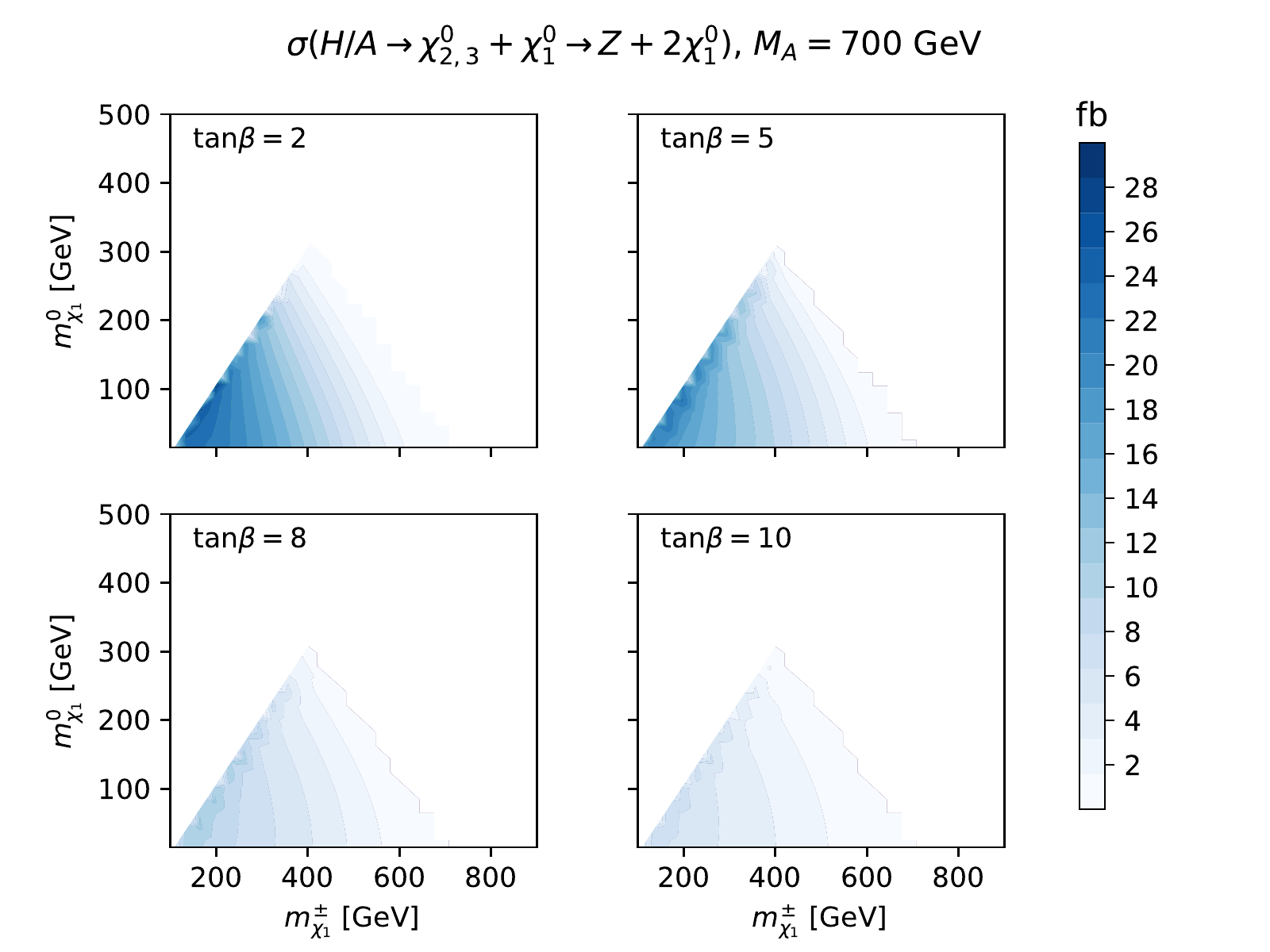} 
\caption{The production cross-section for the process $H/A \to \chi_{2,3}^0 \chi_1^0 \to Z+2\chi_1^0$ for higgsino-like electroweakinos, with heavy scalar mass $m_A = 600, 700$ GeV and $\tan\beta = 2~,5,~8,~10$.}
\label{fig:higgs_xsection-700}
\end{figure}

\begin{figure}[htb]
\centering
	\includegraphics[scale=0.5]{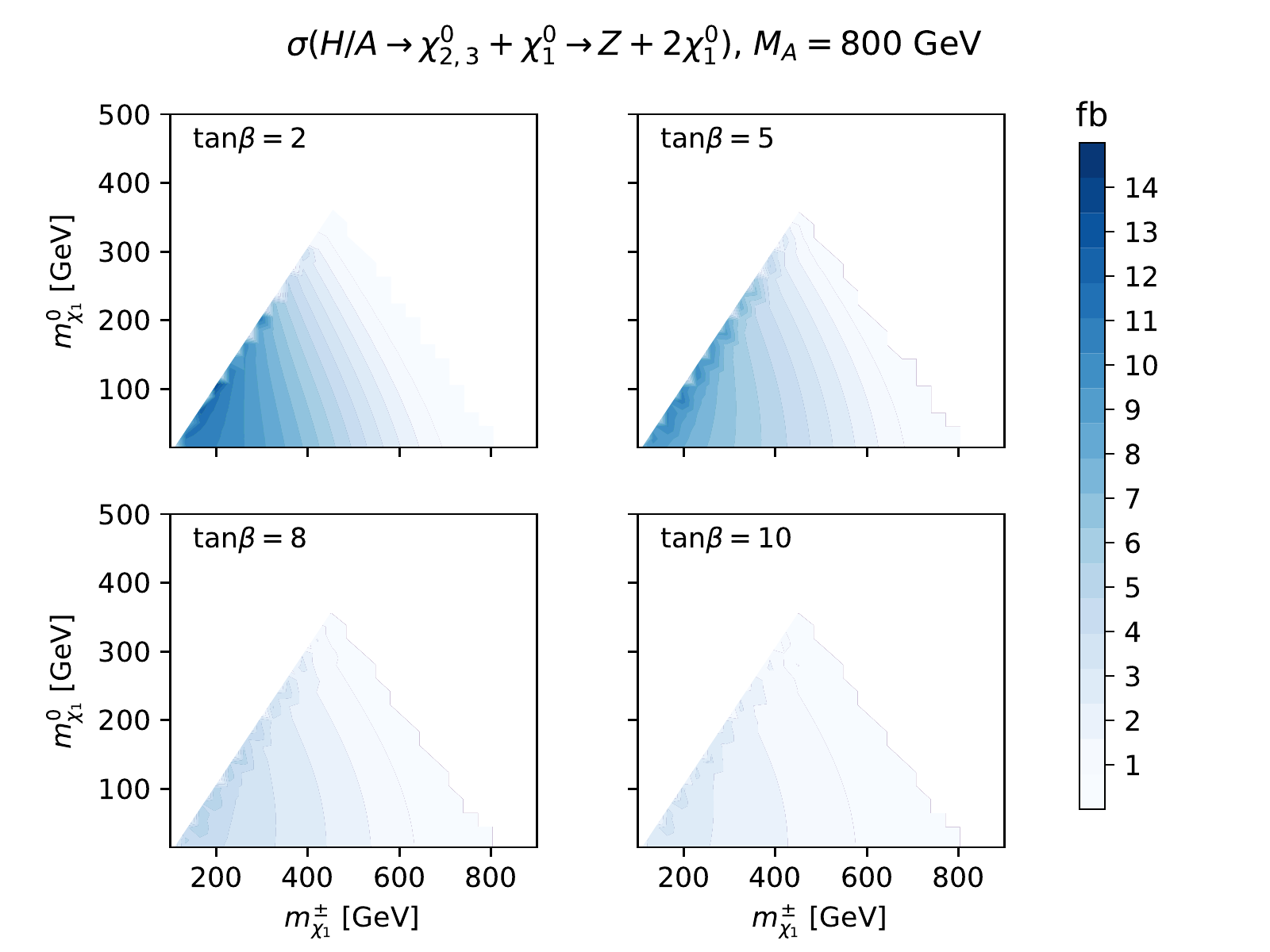} 	\includegraphics[scale=0.5]{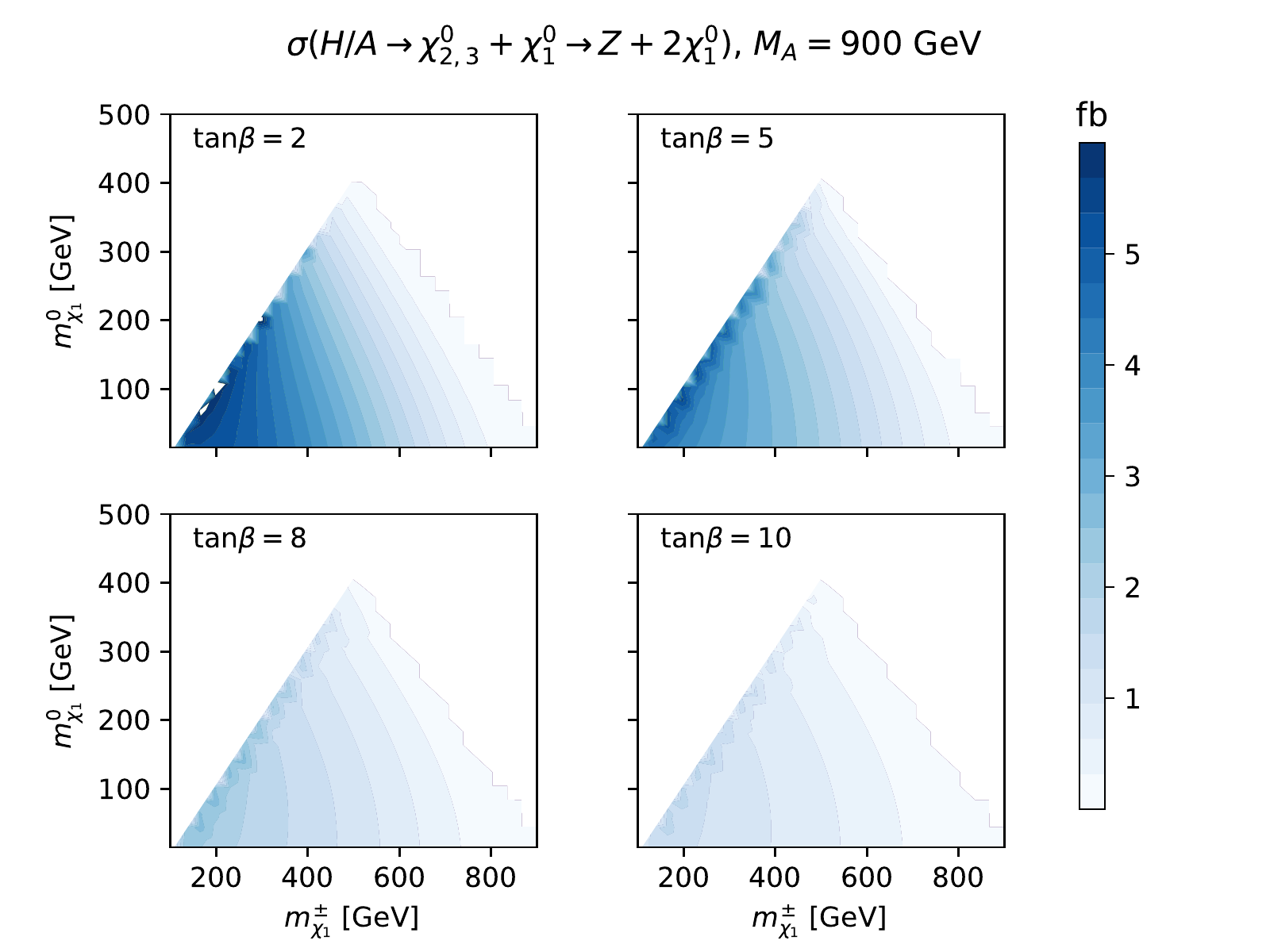} 
\caption{The production cross-section for the process $H/A \to \chi_{2,3}^0 \chi_1^0 \to Z+2\chi_1^0$ for higgsino-like electroweakinos, with heavy scalar mass $m_A = 800, 900$ GeV and $\tan\beta = 2~,5,~8,~10$.}
\label{fig:higgs_xsection-900}
\end{figure}

In Fig.~\ref{fig:higgs_xsection}, we have shown the production cross-section for the process $H/A \to \chi_{2,3}^0 \chi_1^0 \to Z+2\chi_1^0$ for higgsino-like electroweakinos in the case of $m_A = 600$ GeV and $\tan\beta = 2~,5,~8,~10$. For reader's convenience, we also show the comparison of the production cross section for $m_A = 600, ~700\text{, GeV and} ~800,~900$ GeV in 
Fig.~\ref{fig:higgs_xsection-700}, and Fig.~\ref{fig:higgs_xsection-900} respectively.

\section{Results for sgn$(\mu)=+1$}

\begin{figure}[htb]
	\centering
	\includegraphics[scale=0.6]{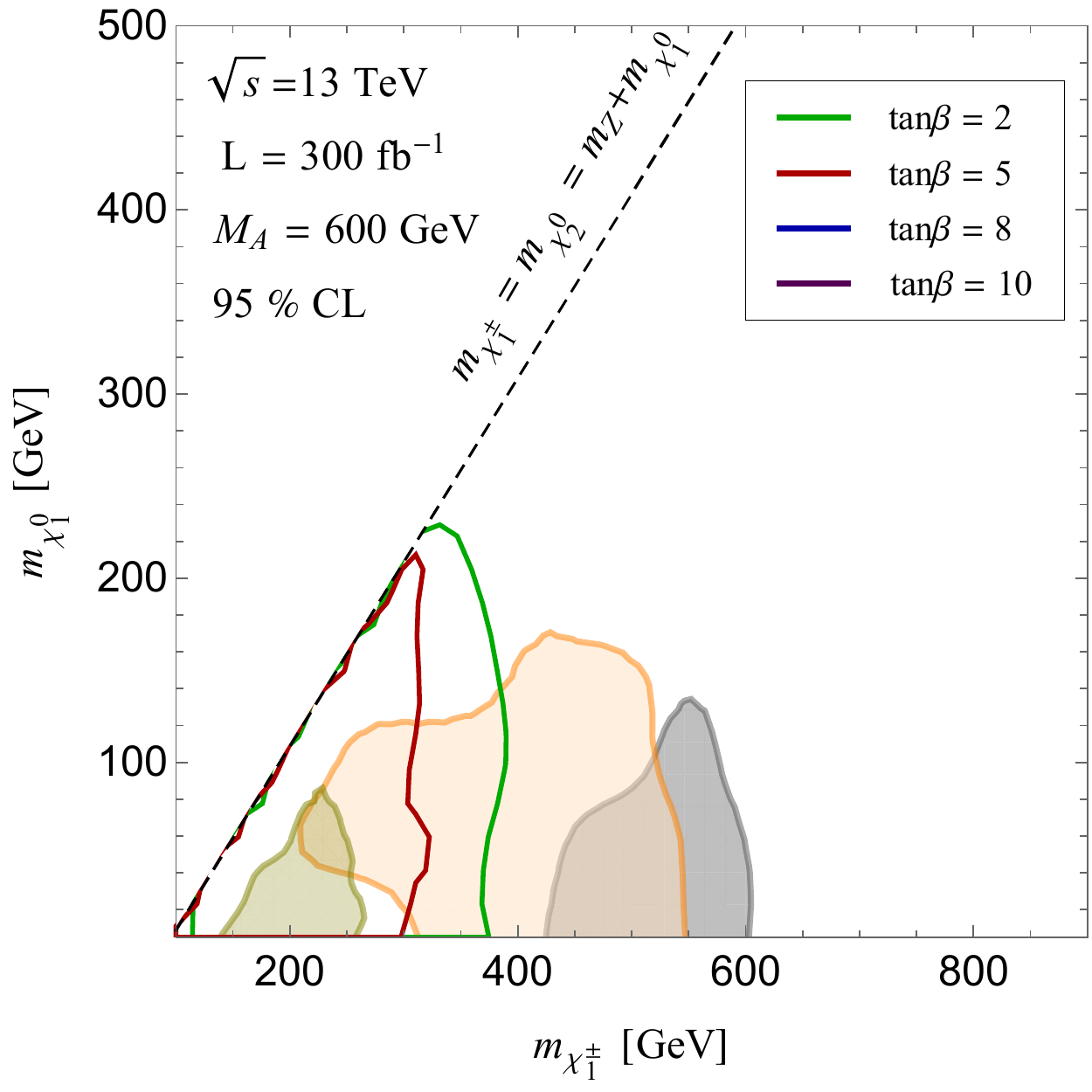}
	\includegraphics[scale=0.6]{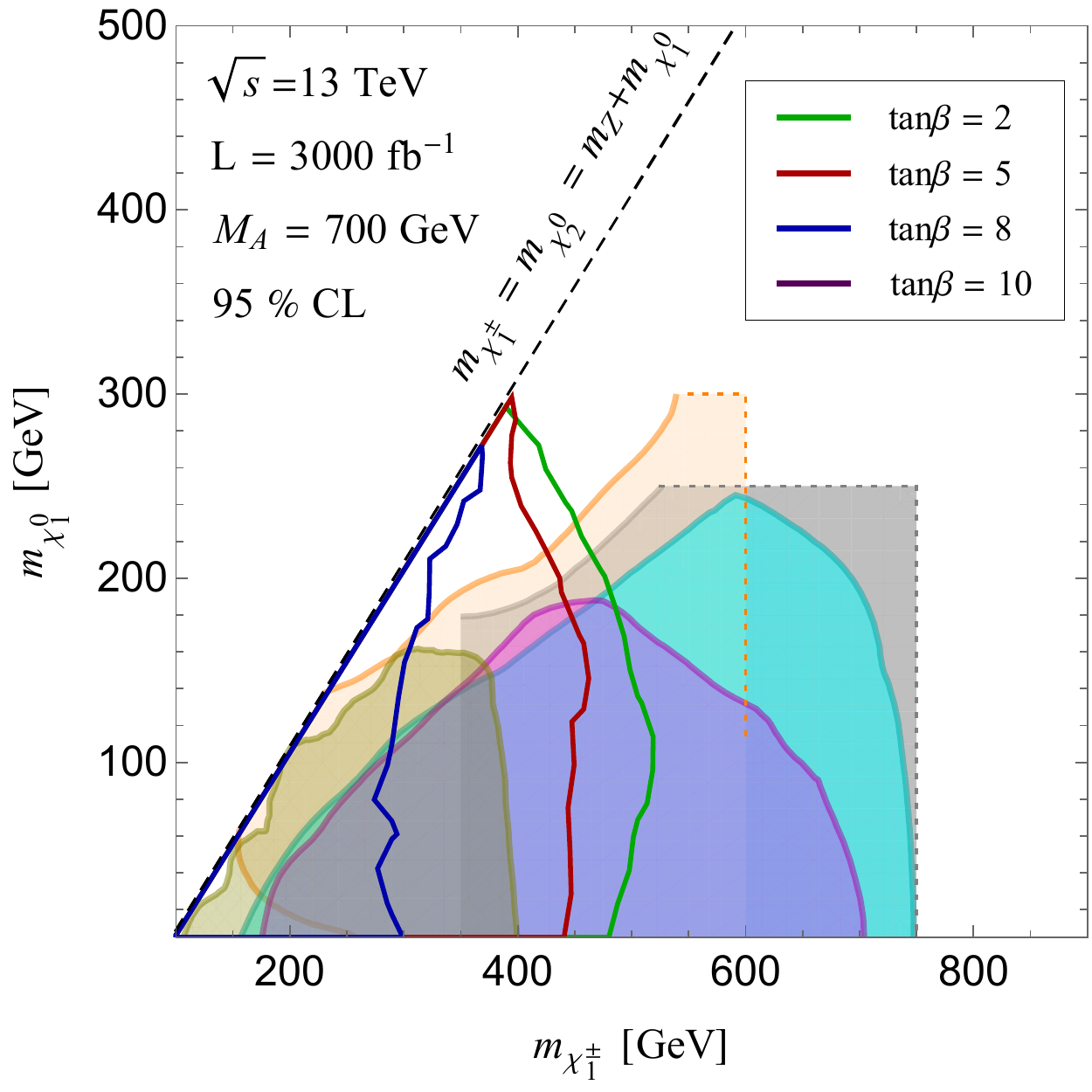}
	\caption{Comparison between the $H/A \to \chi_h^0 \chi_1^0 \to Z+2\chi_1^0$ electroweakino search and the existing direct production constraints for the higgsino-like electroweakinos at HL-LHC assuming sgn$(\mu)=+1$. In the case of the resonant Higgs channel we present the bounds for $m_A = 600 ~(700)$ GeV for luminosity of $300 ~(3000) \text{ fb}^{-1}$.}
	\label{fig:higgs_bounds_mu_plus}
\end{figure}

In the main results presented in this paper we have assumed that sgn$(\mu)=-1$. Apart from the motivations given in the introduction for this choice, it is of interest to also explore the possibility of sgn$(\mu)=+1$. We have studied the corresponding production cross sections in this case and find only minor qualitative differences to the main results. However, for completeness in Fig. \ref{fig:higgs_bounds_mu_plus} we show the comparison between $H/A \to \chi_h^0 \chi_1^0 \to Z+2\chi_1^0$ electroweakino search and the existing direct production constraints for the higgsino-like electroweakinos at HL-LHC assuming sgn$(\mu)=+1$. In the case of the resonant Higgs channel we present the bounds for $m_A = 600 ~(700)$ GeV for $300 ~(3000)\text{ fb}^{-1}$ luminosity.

\bibliography{ref}

\bibliographystyle{JHEP}

%%%%%%%%%%%%%%%%%%%%%%%%%%%%%%%%%%%%%%%%%%%%%%%%%%%%%%%%%
\end{document}